\documentclass[isre,nonblindrev]{informs4}

\usepackage{eqndefns-left}
\Equationvalidatefalse
\RequirePackage{tgtermes}
\RequirePackage{newtxtext}
\RequirePackage{newtxmath}
\RequirePackage{bm}
\RequirePackage{endnotes}

\OneAndAHalfSpacedXII

\usepackage{multirow}
\usepackage{graphicx}
\usepackage{amsmath}
\usepackage{float}
\usepackage{placeins}

\makeatletter
\def\Dskips#1#2#3#4{\abovedisplayskip 4\p@ \@plus2\p@ \@minus2\p@
   \abovedisplayshortskip \z@ \@plus2\p@
   \belowdisplayshortskip 4\p@ \@plus2\p@ \@minus2\p@
   \belowdisplayskip \abovedisplayskip}%
\makeatother

\usepackage{natbib}
 \bibpunct[, ]{(}{)}{,}{a}{}{,}%
 \def\bibfont{\small}%

\EquationsNumberedThrough
\TheoremsNumberedThrough
\ECRepeatTheorems

\theoremstyle{TH}
\newtheorem{axiom}{Axiom}
\theoremstyle{THkey}
\newtheorem{repeataxiom}{Axiom}

\newcommand{\Dom}{\mathcal{D}}
\newcommand{\Rplus}{\mathbb{R}_{+}}
\newcommand{\eb}[1]{\mathbf{e}_{#1}}
\newcommand{\sv}[1]{\mathbf{s}_{#1}}
\newcommand{\Hut}{H_{ut}}
\newcommand{\gs}{g^{*}}
\newcommand{\Gs}{\mathcal{G}^{*}}

\makeatletter
\def\theARTICLEABSTRACT{%
  \HOOKb
  \vspace*{18pt}
  \begin{minipage}[t]{\textwidth}\parindent1em
    \ABSfont
    \noindent\theABSTRACT\endgraf
    \vskip5pt
    \theFUNDING
    \theKEYWORDS
    \theSUBJECTCLASS
    \theAREAOFREVIEW
    \theMSCCLASS
    \theORMSCLASS
    \if@BLINDREV\else\theHISTORY\fi
    \noindent\hrulefill
  \end{minipage}%
  \vspace*{0pt}
}
\makeatother

\MANUSCRIPTNO{}

\makeatletter
\long\def\theARTICLETOPLEFT{}
\def\theARTICLETOPRIGHT{}
\makeatother
\long\def\ARTICLEAUTHORS#1{\gdef\theARTICLEAUTHORS{\begin{center}\HOOKa\vspace*{0pt}#1\vspace*{0pt}\end{center}}}
\def\AUTHOR#1{\begin{center}\AUTHORfont\HD{15}{0}#1\HD{0}{6}\relax\end{center}}
\def\AFF#1{\begin{center}\AFFfont{#1}\relax\vskip1.6pt\end{center}}
\RRHSecondLine{\fs.7.9.\rm Working paper}
\LRHSecondLine{\fs.7.9.\rm Working paper}
\begin{document}

\RUNAUTHOR{Li and Tuzhilin}

\RUNTITLE{When Does a Recommendation Spark Curiosity?}

\TITLE{When Does a Recommendation Spark Curiosity? Theory, Measurement, and Field Evidence}

\ARTICLEAUTHORS{%
\AUTHOR{Pan Li}
\AFF{Scheller College of Business, Georgia Tech, \EMAIL{pan.li@scheller.gatech.edu}}
\AUTHOR{Alexander Tuzhilin}
\AFF{Stern School of Business, New York University, \EMAIL{atuzhili@stern.nyu.edu}}
}

\ABSTRACT{%
Consumers on digital platforms are driven not only by what they like, but also by what they are curious about. Curiosity, the drive to close the gap between what one knows and what one wants to know, pulls consumers beyond their familiar content, sustains engagement, and breaks them out of the filter bubbles of repetitive, relevance-only recommendations. Despite its importance, curiosity is hardly incorporated into modern recommender systems for lack of a formal measure. The information-gap theory of curiosity provides no formal quantitative model that a recommender system can operate on, and existing proxy metrics (e.g., unexpectedness, variety-seeking, and Wundt-curve arousal) do not accurately measure the curiosity that a recommended item (e.g., a product, movie, or video) evokes. To fill this research gap, we formalize the information-gap theory into five axioms of curiosity and construct the canonical formulation of a computational measure of the curiosity that a recommended item evokes in a consumer. We validate our curiosity measure through a survey-based study, in which it matches consumers' self-reported curiosity and outperforms alternative proxy metrics. We subsequently build a curiosity-aware recommender system that incorporates this measure into the utility function design, and we show through offline experiments on three real-world datasets that it significantly outperforms the state-of-the-art methods. Furthermore, we conduct a large-scale online controlled experiment at a major video-streaming platform in Asia, where we show that our method significantly increases click-through rate, video views, and time spent over the latest production system, leading to tangible economic impact for the company. Our method has since been deployed on the platform and remains in operation. Our findings establish consumer curiosity as a measurable and actionable objective for recommender systems, with important managerial implications for digital platforms to recommend what consumers wonder about.
}%

\KEYWORDS{Curiosity; Information-Gap Theory; Recommender Systems; Computational Measure; Online Experiment}

\maketitle

\raggedbottom

\section{Introduction}\label{sec:intro}

Curiosity is an important motivation that drives consumers to close an information gap between what they actually know and what they want to know \citep{loewenstein1994,menon2002}. In particular, curiosity shapes how consumers evaluate the products they encounter \citep{petty2001}, how much they consume \citep{raju1980,roseman1990}, and how willing they are to explore beyond what they already know \citep{berlyne1966}. Therefore, marketers have extensively utilized this concept in practice, for example, by withholding an advertisement's sponsor until the end of the commercial \citep{pavelchak1988} and by writing headlines that hold back the key fact of a story so that readers would click on the link to find it out \citep{scacco2020}.

Recommender systems, meanwhile, are a natural setting for using consumer curiosity since many existing methods repeatedly recommend to consumers products similar to what they have already consumed \citep{kaminskas2016}, resulting in filter bubbles \citep{pariser2011}, reduced sales diversity \citep{fleder2009}, and consumer dissatisfaction \citep{kapoor2015}. In contrast, a consumer nudged toward an information gap between what she/he knows about a product and what she/he wants to know will be curious about the product and, therefore, will be more likely to accept the recommendation and remain engaged with the system. However, incorporating curiosity into recommendations is a highly non-trivial task for the following three reasons.

First, the information-gap theory \citep{loewenstein1994}, which has become the foundation of the subsequent literature on curiosity, identifies the key theoretical constructs relevant to curiosity. However, it provides no formal quantitative model, functional form, or measurement that a recommender system could operationalize directly. 

Second, the proxy metrics that do exist in the recommender system literature, such as unexpectedness \citep{adamopoulos2014}, variety-seeking \citep{li2024}, and Wundt-curve arousal \citep{fu2024}, do not capture how the gap that evokes the most curiosity depends on what the consumer already knows. To illustrate this point, consider two items (i.e., products or content that a recommender system can recommend) at the same distance from the consumer's overall taste, where the consumer has already consumed many items similar to the first one but none similar to the second one. These proxy metrics score the two items identically, while curiosity requires \emph{partial knowledge}, so the first item opens a gap that the consumer can perceive, and the second one opens none.

Third, there is no validated procedure for computing consumer curiosity from consumption records at the scale of an industrial platform, since none of the existing exploration-oriented recommender systems \citep{zhao2016,fu2024} has been evaluated against the curiosity that consumers actually report toward individual items, either in controlled studies or in production settings.

To fill this research gap, namely the lack of a formally defined, validated, and scalable metric of consumer curiosity, we propose a computational measure that quantifies the level of curiosity that a recommended item evokes in a consumer. It is based only on the consumption records of that consumer and, therefore, does not require any explicit consumer feedback. To introduce this measure, we formalize the information-gap theory \citep{loewenstein1994} by proposing a set of five axioms over three components of the consumer's knowledge state, namely (a) the \emph{information gap} between the consumer and the item, (b) the \emph{surrounding context} that the consumer has accumulated around the item, and (c) the \emph{reference gap} that the consumer has adapted to over time. We prove that these five axioms, together with additional shape conditions, determine a canonical functional form of curiosity. Furthermore, we show how to operationalize each component of the proposed framework within the context of recommender systems, and we validate the measure by comparing it to consumers' self-reported curiosity in a survey-based study.

We subsequently build a curiosity-aware recommender system that incorporates our proposed curiosity measure into the recommendation process, where we design a utility function that combines the curiosity objective with the relevance objective of the underlying recommendation model. Under our proposed framework, any standard relevance model can be augmented with the curiosity objective without redesigning the deployed system. Through offline experiments on three real-world datasets from Yelp, MovieLens, and Alibaba-Youku, we demonstrate that our curiosity-aware recommender system significantly improves recommendation performance over state-of-the-art baseline methods across all the evaluation metrics, which capture the accuracy, diversity, and novelty of the recommendations, and that the improvements hold across a wide range of experiment settings. Moreover, we conducted a large-scale online controlled experiment at a major video-streaming platform in Asia, where our curiosity-aware recommender system significantly increased click-through rate, video views, and time spent over the latest production system. The gains increased with the depth of viewing histories and the recency of activity, which matches the information-gap theory. Following these results, the company has since deployed our proposed method in production.

In this paper, we make the following research contributions. First, we propose a curiosity framework that formalizes the information-gap theory of curiosity with a set of axioms, and we derive the canonical curiosity measure from these axioms. Second, we validate the proposed measure through a survey-based study, in which the curiosity measure matches consumers' self-reported curiosity, and we empirically test each of the five axioms across our experiments. Third, we propose a curiosity-aware recommendation framework that incorporates the curiosity objective into recommendation modeling, and we demonstrate through extensive offline experiments that this approach significantly outperforms state-of-the-art recommendation baselines. Finally, we conducted a large-scale online controlled experiment at a major video-streaming platform, and we demonstrate that our curiosity-aware recommender system significantly increased consumer engagement over the latest production system deployed on that platform, generating tangible business impact. Consequently, this paper presents the first axiomatic measure of consumer curiosity for recommender systems, which is validated against consumers' self-reported curiosity and deployed in production. Collectively, these results establish consumer curiosity as a measurable and actionable objective for recommender systems, leading to important managerial implications for digital platforms.

\section{Literature Review}\label{sec:background}
\vspace{6pt}

\subsection{Curiosity and Information-Gap Theory}\label{sec:infogap}

Consumer curiosity has been studied extensively in psychology and marketing \citep{loewenstein1994,menon2002,han2007}, where it influences both consumers' evaluations of what they encounter \citep{petty2001} and their consumption patterns \citep{lerner2000}. Affect-as-information theory \citep{schwarz1990} explains that consumers treat how a product makes them feel as information about the product itself, so curiosity raises both their evaluation of the experience and their subsequent consumption behaviors \citep{pavelchak1988}. Strategies for triggering curiosity have been developed across social media \citep{scacco2020}, search \citep{ripberger2011}, and advertising \citep{menon2002,daume2020}. 

Among various approaches to defining curiosity, the information-gap framework \citep{loewenstein1994} has received the most attention. It presumes that curiosity arises when a consumer becomes aware of a gap between what she/he actually knows and what she/he wants to know, and that the discomfort of this awareness motivates her/him to close the gap. There are two important components of the theory. First, the gap is defined \emph{relative to a reference point}, so the same objective quantity of missing information may or may not evoke curiosity, depending on where the consumer's reference point sits. Second, curiosity requires a \emph{pre-existing knowledge base}, since a consumer who knows nothing related to an item cannot perceive that anything is missing and feels only indifference.

However, the information-gap theory \citep{loewenstein1994} provides no formal quantitative model that a recommender system can operate on. It is worth noting that a subsequent model formalizes preferences regarding the presence and absence of information gaps \citep{golman2018}, while it does not provide a measure of the curiosity that an item evokes in a consumer. In this paper, we formalize the structural claims of the theory in the form of a set of axioms and derive the canonical form of the curiosity measure that satisfies them.

\subsection{Unexpectedness and Related Concepts in Recommender Systems}\label{sec:notdeviation}

Although curiosity, as characterized by information-gap theory, has not been incorporated into recommender systems, there are some relevant concepts in the computer science literature, including unexpectedness, variety-seeking, and Wundt-curve arousal, which we present below together with their differences from the concept of curiosity.

\emph{Unexpectedness} \citep{adamopoulos2014} scores an item by its deviation from the set the consumer would have expected, through a unimodal function of that distance with a most preferred distance for each consumer. The score, therefore, depends on the distance alone, as two items that sit equally far from the expected set receive the same score, no matter how much or little the consumer knows about each of them. Curiosity, meanwhile, requires surrounding context for a gap to be perceived.

\emph{Variety-seeking} \citep{li2024} treats exploration as a stable consumer-level trait that adjusts the degree of unexpectedness, providing more unexpected recommendations to consumers who seek variety and fewer to those who do not. It answers the question of which consumers want more exploration on average, whereas curiosity addresses the question of which item opens a perceivable information gap for this consumer at this moment. Two consumers with identical variety-seeking scores can differ in their momentary readiness for exploration, and the curiosity level of the same consumer differs from one day to the next as recent consumption shifts her/his reference point.

The \emph{Wundt-curve} arousal model \citep{fu2024} maintains that an item that is barely surprising leaves the consumer unmoved, a very surprising one is unpleasant to the consumer, and the level in between produces the strongest response. That most preferred level is a single value for each consumer that is estimated from her/his past consumption. Curiosity also rises and then falls with the gap, but the gap that maximizes it depends on how much the consumer knows about the particular item and on her/his recent consumption, so it differs from item to item and changes over time.

Besides these three most relevant concepts, there are additional metrics in the recommender systems literature, such as \emph{novelty} \citep{kaminskas2016,zhao2016}, \emph{diversity} \citep{gunawardana2015}, which re-ranking methods including maximal marginal relevance (MMR) \citep{carbonell1998} and determinantal point processes (DPP) \citep{chen2018dpp} optimize, and \emph{serendipity} \citep{kaminskas2016}. We summarize the differences between curiosity and all these related concepts in Table \ref{tab:concepts}.

\begin{table}[tbp]
\TABLE
{Systematic Comparison of Curiosity and Related Concepts\label{tab:concepts}}
{\footnotesize
\begin{tabular}{@{}p{0.1\textwidth}p{0.29\textwidth}p{0.17\textwidth}p{0.17\textwidth}p{0.22\textwidth}@{}}
\hline\up
Concept & What it measures & Level & Reference point & Response shape \\ \hline\up
Novelty & Whether the item is new or unfamiliar to the consumer & Consumer--item & Consumption history & Monotone increasing in distance \\[3pt]
Diversity & Dissimilarity among the items within one recommendation list & Item list & The list itself & Increasing in pairwise distance \\[3pt]
Serendipity & Pleasant surprise: unexpectedness combined with relevance & Consumer--item & Expected set & Monotone in deviation, gated by relevance \\[3pt]
Unexpectedness & Distance of the item from the consumer's expected set & Consumer--item & Expected set & Inverted-U with a peak fixed per consumer \\[3pt]
Variety-seeking & Stable propensity to seek different content that scales unexpectedness & Consumer (trait) & Consumption history & Monotone, scaled by the trait \\[3pt]
Wundt-curve arousal & Arousal response to surprise with one optimal stimulus level & Consumer--item & Surprise stimulus axis & Inverted-U with a fixed peak \\[3pt] \hline
\textbf{Curiosity} & Curiosity evoked by an information gap that the consumer can perceive & Consumer--item--time (state) & Reference gap that adapts to recent consumption & Inverted-U whose peak shifts with the surrounding context and the reference gap \down\\ \hline
\end{tabular}}
{}
\end{table}

\subsection{Curiosity-Driven Exploration in Machine Learning}\label{sec:machinecuriosity}

The concept of curiosity has also been used in machine learning \citep{wu2013}, where it refers to the learning system and not to the consumer. For example, in curiosity-driven reinforcement learning, an agent receives an intrinsic reward for visiting the states of the environment that it cannot yet predict, formalized as the agent's prediction error \citep{pathak2017} or learning progress \citep{schmidhuber1991,oudeyer2007}. Therefore, the agent explores the regions it understands least and is drawn to close its own knowledge gaps, just as a curious consumer is drawn to close her/his own. Recommender systems implement the same idea on the platform side through contextual bandits \citep{li2010}, where the system occasionally recommends items with few responses, trading some immediate engagement for faster learning of consumer preferences.

These machine-learning ``curiosity'' approaches differ from ours in whose curiosity is being modeled. Machine curiosity and bandit exploration quantify the \emph{system's uncertainty} about the consumer and reward the system for reducing it, so an item is explored because the platform knows little about the consumer's response to it. In contrast, curiosity quantifies the \emph{consumer's information gap} about a particular item. In that sense, an item is recommended because the consumer perceives something to learn from it. The two concepts, therefore, are complementary to each other. A bandit determines when the platform should explore to improve its own estimates, whereas curiosity identifies which items the consumer wants to explore, and our proposed curiosity measure can be incorporated into any recommender system regardless of whether or not that system explores.

\section{Curiosity Framework}\label{sec:construct}
\vspace{6pt}

\subsection{Preliminaries}\label{sec:prelim}

Let $U$ denote the set of consumers, $I$ the set of items, and $T$ the set of discrete time periods, and let $\Rplus$ denote the set of non-negative real numbers. For each consumer $u$ and time $t$, $\Hut$ is the consumer's \textit{consumption history}, recorded as the set of consumption events $(j,t')$ with $t' < t$, and we write $j \in \Hut$ when item $j$ appears in an event of $\Hut$. An item enters the history when the consumer rates or watches it, and we treat consumption as binary and permanent. We embed both the items and consumers in a shared metric space $(E,d)$, where the \textit{item vector} $\eb{i} \in E$ represents the information that item $i$ carries, which is the same for every consumer and stays fixed over the modeling window. The \textit{consumer state} $\sv{ut} \in E$ represents what consumer $u$ knows at time $t$, which is determined by summarizing the consumption history with a sequential encoder $\mathrm{Enc}(\cdot)$, i.e., $\sv{ut} = \mathrm{Enc}(\Hut)$, and $\sv{ut}$ is updated whenever the consumer consumes a new item. By placing both representations in the same space, we can compute the distance $d(\eb{i},\sv{ut})$ directly between an item and a consumer. We summarize all relevant notation in Table \ref{tab:notation}.

\begin{table}[tbp]
\TABLE
{Notation\label{tab:notation}}
{\footnotesize
\begin{tabular}{@{}p{0.24\textwidth}p{0.68\textwidth}@{}}
\hline\up
Symbol & Meaning \\ \hline\up
$U$, $I$, $T$ & Sets of consumers, items, and discrete time periods \\[3pt]
$\Hut$ & Consumption history, the consumption events $(j,t')$ of consumer $u$ before time $t$ \\[3pt]
$(E,d)$ & Representation space $E \subseteq \mathbb{R}^{p}$ of dimension $p$ and its metric $d$ \\[3pt]
$\eb{i}$, $\sv{ut}$ & Item vector and consumer state, with $\sv{ut} = \mathrm{Enc}(\Hut)$ for a sequential encoder $\mathrm{Enc}(\cdot)$ \\[3pt]
$g_{uit}$ & Information gap, $d(\eb{i},\sv{ut})$ \\[3pt]
$f_{uit}$ & Surrounding context, the acquired knowledge lying near item $i$ \\[3pt]
$K$, $\Lambda$, $\rho$ & Kernel, saturating map, and bandwidth that define $f$ (Section \ref{sec:context}) \\[3pt]
$\tau_{ut}$ & Reference gap, the gap level the consumer has adapted to \\[3pt]
$\omega_{jt'}$, $\lambda$ & Recency weights and decay parameter that define $\tau$ (Section \ref{sec:refgap}) \\[3pt]
$C$, $\Dom$ & Curiosity measure and its domain $\{(u,i,t) : i \notin \Hut\}$ \\[3pt]
$F$ & Functional form of the measure, $C(u,i,t) = F(g_{uit},f_{uit},\tau_{ut})$ (Axiom \ref{ax:relativity}) \\[3pt]
$\Gs(f,\tau)$, $\gs(f,\tau)$ & Set of curiosity-maximizing gaps and its largest element \\[3pt]
$\sigma^{2}$ & Width of the canonical measure of Theorem \ref{thm:char} \\[3pt]
$A(f)$, $m(f,\tau)$, $\psi$ & Gate, peak location, and profile in the shape conditions (S1)--(S4) \down\\ \hline
\end{tabular}}
{}
\end{table}

Note that the information-gap theory \citep{loewenstein1994} identifies three important theoretical concepts relevant to curiosity: (1) \emph{information gap}, as curiosity comes from a gap between what the consumer knows and what she/he wants to know; (2) \emph{surrounding context}, as the consumer needs knowledge already in place around the item to perceive that anything is missing at all; (3) \emph{reference gap}, as the gap is felt relative to a reference point that adapts as the consumer learns. We now introduce these three components below, and we present their formulation details in Section~\ref{sec:measurement}.

\begin{itemize}
\item \textbf{Information Gap} $g_{uit} = d\left( \eb{i},\sv{ut} \right) \in \Rplus$ is the distance from consumer $u$'s current knowledge $\sv{ut}$ to item $i$ and represents the gap between what she/he knows and what she/he wants to know. It measures how much the consumer has yet to learn about that item.
\item \textbf{Surrounding Context} $f_{uit} = \Lambda\!\left( \sum_{j \in \Hut} K(\eb{j},\eb{i}) \right) \in [0,1]$ measures how much of what the consumer knows lies near item $i$. Here, the kernel $K$ is a similarity function that converts the distance $d(\eb{j},\eb{i})$ between two items into a weight, largest when the two items coincide and decaying toward zero as the distance grows, and $\Lambda$ maps the accumulated weight onto $[0,1]$. This component captures the requirement that a consumer perceives a gap about an item only when she/he already knows something about related items.
\item \textbf{Reference Gap} $\tau_{ut} = \sum_{(j,t') \in \Hut} \omega_{jt'}\, g_{ujt'} > 0$ is the gap level that the consumer has grown used to, and it serves as the adaptive reference point against which a gap is felt. $g_{ujt'}$ is the information gap that item $j$ opened at time $t'$, and the weight $\omega_{jt'}$ \textit{decays} with the elapsed time $t - t'$, so that the recently consumed items count more.
\end{itemize}

Based on these three components, we define a \textit{curiosity function} as a continuous mapping
\[
C : \Dom \rightarrow \Rplus,
\qquad
\Dom = \{(u,i,t) : i \notin \Hut\},
\]
where $C(u,i,t)$ is the curiosity that item $i$ evokes in consumer $u$ at time $t$. We would like to highlight two properties of this function. First, curiosity $C$ is a \emph{state} of a consumer regarding an item at one particular moment, and it can change from moment to moment. Second, $C$ is defined only for the items that the consumer has not consumed, since an item whose information she/he has already absorbed leaves no gap to close and therefore evokes no curiosity, so we exclude these items from the domain $\Dom$, consistent with the \emph{closure restriction} \citep{loewenstein1994}. Next, we state the five axioms of curiosity, each of which is a structural claim of information-gap theory.

\subsection{The Five Axioms of Curiosity}\label{sec:axioms}

\begin{axiom}[Gap Relativity]\label{ax:relativity}
There exists a continuous function $F : \Rplus \times [0,1] \times (0,\infty) \rightarrow \Rplus$ such that $C(u,i,t) = F\left( g_{uit},f_{uit},\tau_{ut} \right)$ for all $(u,i,t) \in \Dom$; i.e., curiosity depends on $(u,i,t)$ only through the gap-relative quantities $(g,f,\tau)$.
\end{axiom}

This axiom restates the defining claim of information-gap theory that \emph{``an information gap can be defined by two quantities: what one knows and what one wants to know''} \citep{loewenstein1994}, and it adds the surrounding context and the reference gap that determine whether and how strongly the gap is felt. Specifically, any two consumer--item--time triples $(u,i,t)$ and $(u',i',t')$ that share the same $(g,f,\tau)$ will stimulate the same curiosity level $F(g,f,\tau)$, regardless of any other differences between the items or the consumers. We empirically validate this axiom in our survey-based study in Section \ref{sec:labresults}, analysis (c), where items matched on $(g,f,\tau)$ but differing in popularity and framing evoke the same curiosity. Given Axiom \ref{ax:relativity}, we denote by $\gs(f,\tau)$ the largest element of $\arg\max_{g \geq 0} F(g,f,\tau)$ and call it the \textit{curiosity-maximizing gap}, the gap that evokes the most curiosity at a given surrounding context and reference gap.

\begin{axiom}[Perceivability]\label{ax:perceivability}
For all $g,\tau$: $F(g,0,\tau) = 0$; and for each $f \in (0,1)$, $\tau > 0$: $\lim_{g \rightarrow \infty} F(g,f,\tau) = 0$.
\end{axiom}

This axiom states that with no surrounding context ($f = 0$), the consumer has nothing to compare the item with and cannot notice that anything is missing, while with an extremely large gap ($g \rightarrow \infty$), the item is too far from what the consumer knows for her/him to tell what is missing. Both conditions come from the information-gap theory, where curiosity requires a \emph{``preexisting knowledge base''} and \emph{``a failure to appreciate what one does not know would constitute an absolute barrier to curiosity''} \citep{loewenstein1994}. In our survey-based study in Section \ref{sec:labresults}, analysis (d), we build a zero-context scenario to test the first condition directly, and the second condition appears as the decline of curiosity at large gaps.

\begin{axiom}[Context-Dependent Optimum]\label{ax:optimum}
For all $\tau > 0$ and $0 < f < f' < 1$, $\gs(f',\tau) < \gs(f,\tau)$.
\end{axiom}

This axiom states that, at a fixed reference gap $\tau$, the information gap that evokes the most curiosity becomes smaller as the surrounding context grows. The anchor is the observation that \emph{``as people gain knowledge in a particular area \ldots those gaps will become smaller relative to what they already know,''} so that \emph{``people are likely to become progressively more curious about the topics that they know the most about''} \citep{loewenstein1994}. An example is the contrast between experts and novices. A film scholar, with many watched films close to the recommended one and therefore a high $f_{uit}$, is drawn to the fine distinctions between two directors' treatments of the same scene, while a newcomer, with few watched films nearby and a low $f_{uit}$, needs a much larger gap to feel that anything worth knowing is missing. In our survey-based study in Section \ref{sec:labresults}, analysis (e), we test the axiom by locating the curiosity peak separately in the low- and high-context scenarios, and we confirm that the peak is located at a smaller gap under high context.

\begin{axiom}[Reference-Gap Tracking]\label{ax:tracking}
For each $f \in (0,1)$ and $0 < \tau < \tau'$: $\gs(f,\tau) < \gs(f,\tau')$.
\end{axiom}

This axiom states that the higher the consumer's current reference gap, the larger the gap that evokes the most curiosity. The anchor is the role of the reference point in the theory: \emph{``curiosity, in this view, arises when one's informational reference point in a particular domain becomes elevated above one's current level of knowledge''} \citep{loewenstein1994}. An example is the contrast between two viewers who have acquired the same amount of information about a director's films, so they share the same $f_{uit}$. The viewer who has recently been experiencing large information gaps has a high $\tau_{ut}$ and is drawn to the director's most unusual work, while the viewer who has stayed close to what she/he knows has a low $\tau_{ut}$ and is drawn to the film nearest to it. In our survey-based study in Section \ref{sec:labresults}, analysis (f), we test the axiom by comparing where the curiosity peak lies for participants with low, medium, and high reference gaps at the same context level.

\begin{axiom}[Recency Adaptivity]\label{ax:recency}
$\tau_{ut}$ is a recency-weighted average of the encountered information gap levels:
\[
\tau_{ut} = \sum_{(j,t') \in \Hut} \omega_{jt'}\, g_{ujt'},
\]
where $\omega_{jt'} \geq 0$ and $\sum_{(j,t')} \omega_{jt'} = 1$, with weights non-increasing and non-constant in elapsed time.
\end{axiom}

This axiom states that the reference gap is an average of the encountered gap levels, where the recent ones count more. The anchor is that a reference point adapts to recent experience and fades as that experience becomes more distant \citep{helson1964,metcalfe2005}. An example is a viewer who has spent the past week on familiar sitcoms. The gaps she/he encountered were small, so $\tau_{ut}$ is small, and recommending a documentary in an unfamiliar field would be more than she/he wants. After a week of branching into new genres, the gaps she/he has encountered are larger, $\tau_{ut}$ rises, and the same documentary will now be much more appropriate. Because the axiom is about changes over time, we test it in our offline ablation in Section \ref{sec:ablation}, where removing recency from the reference gap lowers the recommendation accuracy on every dataset, and in our online experiment in Section \ref{sec:fieldhet}, where the treatment effect is larger for recently active consumers.

To summarize, each of these five axioms restates one claim of the information-gap theory in terms of the information gap, the surrounding context, and the reference gap, as listed in Table \ref{tab:foundations}. We further discuss the implicit assumptions that each axiom carries in Appendix \ref{app:foundations}. These assumptions define the settings in which the framework applies.

\begin{table}[tbp]
\TABLE
{The Axioms and Their Foundations in Information-Gap Theory\label{tab:foundations}}
{\begin{tabular}{@{}p{0.29\textwidth}p{0.43\textwidth}p{0.22\textwidth}@{}}
\hline\up
Axiom & Claim in information-gap theory & Empirical test \\ \hline\up
\#\ref{ax:relativity} Gap Relativity & Curiosity depends only on $(g,f,\tau)$. & Survey, Section~\ref{sec:labresults}(c) \\[3pt]
\#\ref{ax:perceivability} Perceivability & No curiosity at zero context or at an extremely large gap. & Survey, Section~\ref{sec:labresults}(d) \\[3pt]
\#\ref{ax:optimum} Context-Dependent Optimum & The optimal gap shrinks as context grows. & Survey, Section~\ref{sec:labresults}(e) \\[3pt]
\#\ref{ax:tracking} Reference-Gap Tracking & The optimal gap grows with the reference gap. & Survey, Section~\ref{sec:labresults}(f) \\[3pt]
\#\ref{ax:recency} Recency Adaptivity & The reference gap is a recency-weighted average of encountered gaps. & Offline (Section~\ref{sec:ablation}) and Online (Section~\ref{sec:fieldhet}) \down\\ \hline
\end{tabular}}
{}
\end{table}

\subsection{The Canonical Curiosity Measure}\label{sec:canonical}

We now present the canonical curiosity measure that satisfies all five axioms introduced in Section \ref{sec:axioms}. 

\begin{theorem}[Characterization]\label{thm:char}
Let $C$ satisfy Axioms \ref{ax:relativity}--\ref{ax:recency}. Then $C$ satisfies the shape conditions (S1)--(S4) if and only if there exists a unique $\sigma^{2} > 0$ such that
\begin{equation}\label{eq:canonical}
F(g,f,\tau) \;=\; f \cdot \exp\left( - \frac{(g - \tau(1 - f))^{2}}{2\sigma^{2}} \right).
\end{equation}
\end{theorem}

\begin{figure}[tbp]
\FIGURE
{\includegraphics[width=0.95\linewidth]{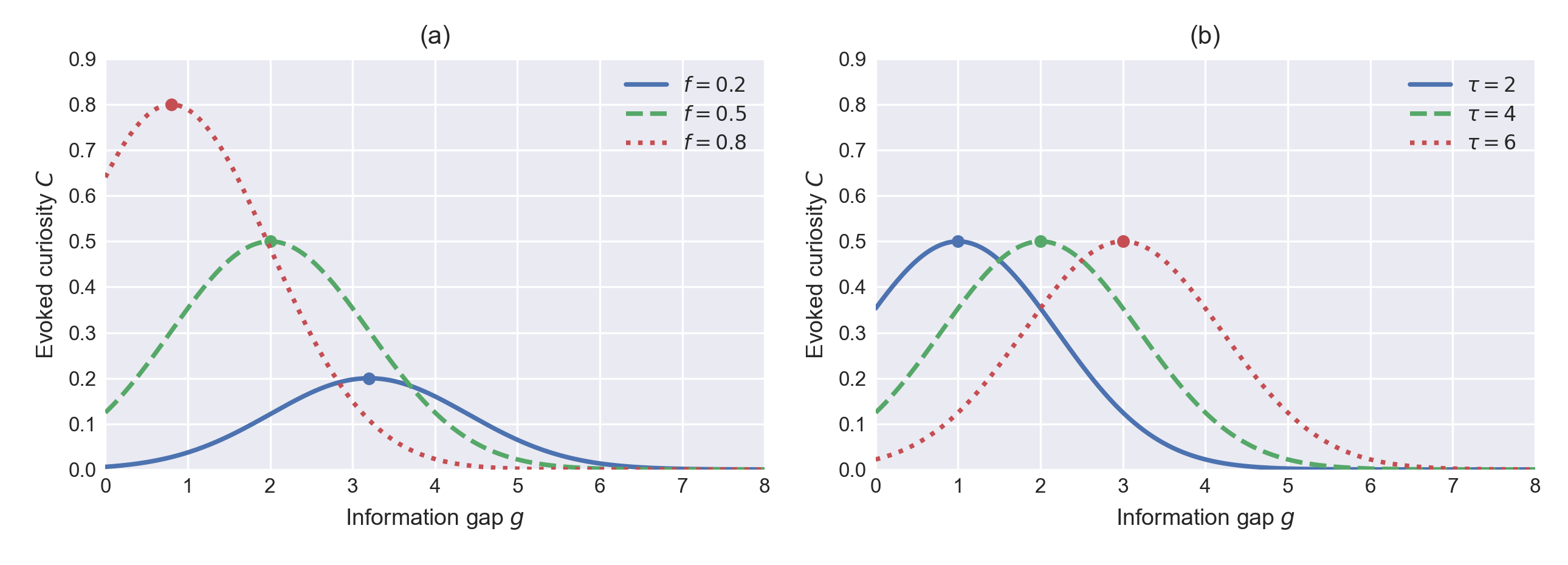}}
{(Color online) Plots of the Canonical Curiosity Measure {\boldmath$F(g,f,\tau) = f \cdot \exp\!\big(-(g - \tau(1 - f))^{2}/2\sigma^{2}\big)$} \label{fig:canonical}}
{(a) Varying the surrounding context $f \in \{0.2, 0.5, 0.8\}$ at $\tau = 4$. (b) Varying the reference gap $\tau \in \{2, 4, 6\}$ at $f = 0.5$. Dots mark the curiosity-maximizing gap $\gs = \tau(1-f)$, and $\sigma = 1.2$ in both panels.}
\end{figure}

The full proof and a discussion of the necessity of the four shape conditions (S1)--(S4) are provided in Appendices~\ref{app:charproof} and~\ref{app:conditions}. The curiosity measure is then $C(u,i,t) = F\left( g_{uit},f_{uit},\tau_{ut} \right)$ for $(u,i,t) \in \Dom$, with $F$ given by \eqref{eq:canonical}. Here, $\sigma^{2}$ determines how quickly curiosity falls as the gap gets farther from the peak. Figure \ref{fig:canonical} presents plots of this canonical measure for different values of $f$ and $\tau$, where curiosity rises and then falls with the gap. The peak shifts toward smaller gaps and rises as the surrounding context $f$ grows in Panel (a), and it shifts toward larger gaps as the reference gap $\tau$ grows in Panel (b).

\subsection{Instantiation of the Canonical Measure}\label{sec:measurement}

As our canonical curiosity measure is constructed over $(g,f,\tau)$, we now turn these three components into computationally measurable quantities in this section.

\subsubsection{The Representation Space and Information Gap $\bm{g_{uit}}$.}\label{sec:repspace}\label{sec:gap}

Both $g$ and $f$ are defined over a metric space $(E,d)$ in which items and consumer states are represented, so we need to specify an item vector $\eb{i}$ and a consumer state $\sv{ut}$ in this metric space.

Two items may differ sharply on explicit features while being near-substitutes in experience, which is an important dimension for modeling the information gap. In movie recommendation, for example, the title, genre, director, cast, and duration of a movie leave out the style and reputation of the director, the popularity of the genre, and the fit of the cast, all of which shape consumer experience \citep{he2017,chen2020}. Therefore, the item vector $\eb{i}$ is not the vector of explicit features itself but a low-dimensional latent vector learned from these features. Specifically, we select an AutoEncoder \citep{kingma2014,hinton2006}, following its successful deployment across marketing applications \citep{timoshenko2019,dhillon2021,dew2022,anand2023} and recommender systems \citep{li2017}, and its scalability in practice \citep{zhang2019,li2020,hardesty2019}. An encoder maps the explicit features to a latent vector and a decoder reconstructs the features from it, with the two trained to minimize the reconstruction error, and the item vector $\eb{i}$ is obtained as the encoder's output. Alternative encoders, such as Item2vec \citep{barkan2016} and Transformer-based models \citep{vaswani2017}, can also be used under our curiosity framework, and we evaluate them in Section \ref{sec:labresults}.

The consumer state $\sv{ut}$ aggregates the consumption history $\Hut$ with the order information, since recently consumed items bear more on the current knowledge state than distant ones \citep{labarbera1983,panniello2016}, so a sequential encoder is appropriate for this task \citep{wang2019}. Among the candidate architectures, including the recurrent neural network (RNN) \citep{rumelhart1986}, long short-term memory (LSTM) \citep{hochreiter1997}, gated recurrent unit (GRU) \citep{cho2014}, and BERT \citep{devlin2019}, we choose GRU, which is widely deployed in recommender systems \citep{hidasi2015,li2020,yu2019} and has been shown to perform well in practice \citep{wang2019}. The GRU processes the consumed items one at a time in the order of consumption and updates its hidden state after each item, and its gates determine how much each consumed item contributes to the updated state. The consumer state $\sv{ut}$ is the hidden state after the GRU has processed the last item consumed before time $t$. To place $\sv{ut}$ in the same space $E$ as the item vectors, we train the GRU so that the distance $d(\eb{i},\sv{ut})$ predicts whether consumer $u$ clicks on item $i$.

Based on the item vector and the consumer state, we can now formulate the information gap as the distance between them, $g_{uit} = d(\eb{i},\sv{ut})$. We use the Euclidean distance in the primary specification for the following reasons. First, the Euclidean distance gives the highest correlation with reported curiosity among the distances that we evaluate in Table \ref{tab:reprgrid}. Second, the surrounding context below accumulates a Gaussian kernel over the same space, so measuring the gap with the same metric keeps $g$ and $f$ on a common scale. Alternative distances, such as the cosine and Mahalanobis distances, can also be used, and we evaluate them in Section \ref{sec:labresults}.

\subsubsection{Surrounding Context $\bm{f_{uit}}$.}\label{sec:context}

Surrounding context measures how much of the consumer's acquired knowledge lies near the item, which is what makes a gap perceivable at all. We measure it as the kernel sum of the consumption history at the item's location, mapped to $[0,1]$:
\[
f_{uit} = \Lambda\!\left( \sum_{j \in \Hut} K_{\rho}(\eb{j},\eb{i}) \right),
\]
where $K_{\rho}(\eb{j},\eb{i}) = \exp\left( -d(\eb{j},\eb{i})^{2}/2\rho^{2} \right)$ is a Gaussian kernel of bandwidth $\rho$. The saturating map $\Lambda(x) = 1 - \exp(-x/x_{0})$ converts this sum into a value in $[0,1)$, where $x_{0}$ is the median kernel sum in the data. Therefore, the first consumed items near item $i$ raise $f_{uit}$ the most, and each further consumed item adds less, so that $f_{uit}$ approaches one as the knowledge around item $i$ accumulates.

The surrounding context $f$ formalizes the requirement of information-gap theory, introduced in Section \ref{sec:prelim} and stated in Axiom \ref{ax:perceivability}, that a consumer notices a gap about an item, i.e., the gap is \emph{perceivable}, only when she/he already knows something related to that item. This requirement leads to the following four properties of $f$. First, $f$ must be \emph{local}, which means that $f_{uit}$ depends on the consumed items that lie close to item $i$ in the representation space and not on the total number of consumed items, since knowledge about unrelated items does not help the consumer notice what is missing about item $i$. Second, $f_{uit}$ must be close to \emph{zero} when none of the consumed items lies near item $i$, since the consumer then has no related knowledge and, by Axiom \ref{ax:perceivability}, feels no curiosity. Third, $f_{uit}$ must \emph{grow} with the number of consumed items near item $i$, since each additional related item adds to the consumer's knowledge around item $i$. Fourth, $f$ must be \emph{graded by proximity}, which means that a consumed item contributes more to $f_{uit}$ the closer it lies to item $i$. The Gaussian kernel has all four properties by construction, since its weight is largest for a consumed item at the location of item $i$, decays toward zero as the distance grows, and adds up over the consumed items. We therefore choose the Gaussian kernel over other kernels under our curiosity framework.


\subsubsection{Reference Gap $\bm{\tau_{ut}}$.}\label{sec:refgap}

Axiom \ref{ax:recency} fixes the form of $\tau_{ut}$ as a convex combination of encountered gap levels with weights non-increasing in elapsed time. We use exponential weights $\omega_{jt'} \propto \lambda^{t-t'}$ in the primary specification, with $\lambda$ tuned on the validation partition of Section \ref{sec:metrics}:
\[
\tau_{ut} = \frac{\sum_{(j,t') \in \Hut} \lambda^{t-t'}\, g_{ujt'}}
                 {\sum_{(j,t') \in \Hut} \lambda^{t-t'}}.
\]
Among the weighting functions that the axiom allows, we choose the exponential decay for three reasons. First, it contains the no-recency benchmark as the limiting case $\lambda = 1$. Second, it fades gradually, as the influence of a previously consumed item shrinks a little more each day, in line with how reference points adapt \citep{helson1964}. Third, it is cheap to compute at scale, since $\tau_{ut}$ can be updated recursively with each newly encountered gap, without storing or rescanning the consumer's history. 


\subsection{Summary of the Curiosity Framework}\label{sec:summary}

To summarize, we measure the curiosity that an item evokes in a consumer with the canonical form of Equation \eqref{eq:canonical}, which follows from the five axioms and the four shape conditions, and we compute its three components from the item vectors and consumer states that a recommender system already learns. Table~\ref{tab:components} summarizes the configuration of each component.

\begin{table}[tbp]
\TABLE
{Configuration of the Components of the Curiosity Measure\label{tab:components}}
{\footnotesize
\begin{tabular}{@{}p{0.21\textwidth}p{0.33\textwidth}p{0.13\textwidth}p{0.27\textwidth}@{}}
\hline\up
Component & Primary specification & Tuned parameter & Alternatives \\ \hline\up
Item vector $\eb{i}$ & AutoEncoder trained on item features & --- & Item2vec, Transformer, tag genome \\[3pt]
Consumer state $\sv{ut}$ & GRU over the consumption history & --- & RNN, LSTM, BERT \\[3pt]
Information gap $g_{uit}$ & Euclidean distance $d(\eb{i},\sv{ut})$ & --- & Cosine, Mahalanobis \\[3pt]
Surrounding context $f_{uit}$ & Gaussian kernel sum mapped to $[0,1]$ by $\Lambda$ & Bandwidth $\rho$ & $k$-nearest neighbors ($k = 5$) \\[3pt]
Reference gap $\tau_{ut}$ & Exponentially weighted average of gaps & Decay $\lambda$ & Sliding window \\[3pt]
Canonical measure $F$ & Equation \eqref{eq:canonical} & Width $\sigma$ & --- \down\\ \hline
\end{tabular}}
{Note: Sections~\ref{sec:labresults} and~\ref{sec:robustness} evaluate the alternatives of every component except the consumer state.}
\end{table}

\section{Curiosity-Aware Recommendation Framework}\label{sec:recsys}

Based on our curiosity measure, we present our curiosity-aware recommendation framework in this section, where we design a utility function that combines the curiosity objective with relevance.

Since curiosity shapes how consumers evaluate an item and how much they consume \citep{petty2001,raju1980,roseman1990}, we incorporate it into the recommendation objective next to the predicted relevance $\hat{r}_{uit}$ of any standard recommendation model. Moreover, consumers differ in how strongly curiosity drives their exploratory behavior \citep{raju1980}, so we weight curiosity by the consumer's \textbf{curiosity responsiveness} $w_{ut} \geq 0$, with $w_{ut} = 0$ meaning that curiosity plays no part in this consumer's engagement. For each consumer, we estimate $w_{ut}$ as the coefficient on $C_{uit}$ in a regression of her/his engagement before time $t$ on $C_{uit}$ that holds predicted relevance fixed, and we constrain the estimate to be non-negative.
Therefore, our recommender system pursues two objectives, as it matches items with consumer preferences and also evokes consumer curiosity, and it solves the following optimization problem:
\[
\max_{i \in \mathcal{C}_{ut}} \; \left( \hat{r}_{uit}, \; w_{ut} \cdot C_{uit} \right)
\]
over the candidate set $\mathcal{C}_{ut}$. An item is Pareto-optimal when no other candidate is at least as good on both objectives and better on at least one, and the platform's policy is a rule that turns the Pareto-optimal set into a ranking. We adopt linear scalarization as the primary design of the proposed utility function $V_{uit}$:
\begin{equation}\label{eq:utility}
V_{uit} = (1-\alpha)\,\hat{r}_{uit} + \alpha\, w_{ut} C_{uit},
\end{equation}
where both terms are standardized, and $V_{uit}$ traces the convex part of the Pareto frontier as $\alpha$ runs from $0$ (pure relevance) to $1$ (pure curiosity). This framing makes $\alpha$ a platform policy parameter that can be determined either manually or through hyperparameter optimization. We study an alternative multiplicative scalarization in Section \ref{sec:robustness}.

Under this design, our curiosity-aware recommender system is particularly flexible, since relevance $\hat{r}$ is exogenous and any standard recommendation model is compatible, including collaborative filtering (CF) \citep{hu2008}, matrix factorization (MF) \citep{koren2009}, neural collaborative filtering (NCF) \citep{he2017}, DeepFM \citep{guo2017}, and the deep interest network (DIN) \citep{zhou2018}, without the need to redesign a deployed system.

\section{Validation of the Curiosity Measure: A Survey-Based Study}\label{sec:labstudy}

To validate the proposed curiosity measure, we conducted a survey-based study in which participants reported the level of curiosity they feel toward each recommended item, and we compare the computed measure $C$ against these self-reports. Following existing designs \citep{menon2002,wang2018}, our study has two goals. First, we establish the \emph{discriminant} validity of $C$ \citep{campbell1959,fornell1981}, as $C$ captures what novelty and unexpectedness cannot and predicts reported curiosity even after we control for both. Second, we test the predictions of the axioms directly on the curiosity that consumers report, which the offline and online experiments cannot observe, since they record only clicks and viewing behavior.

\subsection{Questionnaire Design}\label{sec:labdesign}

\paragraph{Participants and procedure.}
Adult participants were recruited and completed the questionnaire online, following standard practice for online consumer research \citep{goodman2017}. The study was approved by the authors' institutional review board. All recommended movies were drawn from the MovieLens dataset \citep{harper2015}. The full questionnaire is presented in Appendix~\ref{app:questionnaire}, and it consists of the following four parts.
\begin{itemize}
\item \textbf{Background questions.} After consenting, each participant reported her/his gender and age, rated how often she/he watches each of ten movie genres, which we use in the stimulus design below, and answered three questions about her/his movie-watching behavior in the past month (e.g., ``In the past month, to what extent did you seek out films outside your comfort zone?''), which together serve as a proxy for the participant's reference gap. The participant was also asked to name one movie that she/he had already seen and loved, which we use in the closure question below.
\item \textbf{Movie recommendations.} The questionnaire then presented fifteen movie recommendations one at a time and in random order. Each recommendation showed the participant the poster, title, genre, and a short textual description of the movie, and Figure~\ref{fig:questionnaire} in Appendix~\ref{app:questionnaire} shows an example.
\item \textbf{Questions about each recommendation.} After each recommendation, the participant answered six questions on a seven-star scale (1 = not at all, 7 = extremely), following the standard practice in consumer research \citep{churchill1979,wang2018}. Three questions measure curiosity (e.g., ``At the current moment, how curious are you about this movie recommendation?''), and we average them into a single curiosity score (Cronbach's alpha between 0.83 and 0.94). Two questions measure perceived novelty and perceived unexpectedness, following their established operationalizations \citep{zhao2016,adamopoulos2014}. The sixth question measures the participant's prior knowledge of the movie (``How much do you already know about this movie?'').
\item \textbf{Closure question and attention check.} After the fifteen recommendations, the questionnaire asked the participant to imagine that we recommended another movie ``chosen to be as similar as possible to'' the movie she/he had named as seen and loved, with the title of that movie shown in the question (i.e., with the smallest information gap we can create), and to answer the same six questions about this recommendation. The questionnaire ended with an instructed-response attention check \citep{oppenheimer2009}.
\end{itemize}

\paragraph{Stratified stimulus design.}
The canonical measure of Theorem \ref{thm:char} implies that, at a given surrounding context $f$ and reference gap $\tau$, curiosity first rises and then falls as the information gap $g$ grows, with its peak at $\gs = \tau(1-f)$, and that this peak shifts to a smaller gap when $f$ is higher, as Axiom \ref{ax:optimum} states and Figure \ref{fig:canonical} shows. To test that the peak shifts, we need to observe curiosity across the whole range of gaps at a low context and again at a high context, where low and high context refer to low and high values of the surrounding context $f_{uit}$ of Section \ref{sec:context}. Randomly chosen movies do not guarantee that each participant sees movies across the whole range of gaps at both context levels, and without such coverage we cannot locate the peak at each context level. Therefore, we choose the movies to fill a small grid of six gap sizes at each of the two context levels, and we add a set of movies with almost no surrounding context, where Axiom \ref{ax:perceivability} predicts that curiosity stays near zero. Every participant rated all of them.

In total, 300 participants completed the questionnaire. Of these, 283 (94.3\%) passed the instructed-response attention check and were retained (M$_{\text{age}}$ = 31.6, SD = 7.9, 43\% women, 51\% men, 4\% non-binary or another identity, 2\% preferring not to say), and the remaining 17 were excluded. This sample yields 4{,}245 film-level observations, which we analyze in Section \ref{sec:labresults}.

\subsection{Analysis and Results}\label{sec:labresults}

For every film--participant pair, we compute the canonical measure $C_{uit}$ of \eqref{eq:canonical}, and subsequently conduct six groups of analyses. In the first two analyses, we show that $C$ measures curiosity and captures what proxy metrics cannot, and in the remaining four, we test the predictions of the axioms.

\paragraph{(a) Construct validity.}
We first compute the Pearson correlation between $C_{uit}$ and the participant's curiosity score. The canonical measure reaches a Pearson correlation of $r = 0.636$ with the designed gap levels and $0.598$ with the computed gaps (both $p < 0.001$). As a comparison, the proxy metrics of unexpectedness \citep{adamopoulos2014}, variety-seeking \citep{li2024}, and Wundt-curve arousal \citep{fu2024} only achieve Pearson correlations of 0.603, 0.589, and 0.611, respectively, which are below the 0.636 of our measure with the designed gap levels. We further recompute $C$ under six combinations of item vector and distance, namely the AutoEncoder vectors with the Euclidean, cosine, and Mahalanobis distances, and the Item2vec, Transformer, and tag-genome vectors with the Euclidean distance, and Table \ref{tab:reprgrid} reports the correlation of each version with reported curiosity. We find that the item vectors of the original Transformer \citep{vaswani2017} reach a correlation of 0.621, which is higher than that of the primary specification and of every proxy metric, which shows that the validity of the measure carries over to the Transformer item vectors.

\begin{table}[tbp]
\TABLE
{Construct Validity Across Representation and Distance Choices\label{tab:reprgrid}}
{\begin{tabular}{@{}lllc@{}}
\hline\up
Item vector & Consumer state & Distance & \shortstack{Correlation\\with reported} \\ \hline\up
AutoEncoder & GRU  & Euclidean & 0.598 \\
AutoEncoder & GRU  & Cosine    & 0.427 \\
AutoEncoder & GRU  & Mahalanobis & 0.251 \\
Item2vec    & GRU  & Euclidean & 0.508 \\
Transformer & GRU  & Euclidean & 0.621      \\
Tag genome  & GRU  & Euclidean & 0.299 \down\\ \hline
\end{tabular}}
{Note: The primary specification is AutoEncoder--GRU--Euclidean.}
\end{table}

\paragraph{(b) Discriminant validity.}
Following the standard rule that a new construct must predict its criterion beyond established related constructs \citep{campbell1959,fornell1981}, we test whether $C$ predicts reported curiosity after we control for perceived novelty and perceived unexpectedness, and we regress reported curiosity on $C$, perceived novelty, and perceived unexpectedness,
\[
\text{Curiosity}_{uit} = \beta_{0} + \beta_{C}\,C_{uit} + \beta_{N}\,\text{Novelty}_{uit} + \beta_{U}\,\text{Unexp}_{uit} + \eta_{u} + \varepsilon_{uit},
\]
with participant fixed effects $\eta_{u}$ and error term $\varepsilon_{uit}$. We find that $\beta_{C}$ remains positive and significant after we control for perceived novelty and perceived unexpectedness ($\beta_{C} = 4.29$, $p < 0.001$). The controls collapse once $C$ is present, as novelty's coefficient is $-0.016$ and unexpectedness's is $-0.042$, both slightly negative and insignificant. For each measure, Table \ref{tab:study1} reports the correlation with reported curiosity, the partial correlations after controlling for perceived novelty and for perceived unexpectedness, and the increase in $R^{2}$ when the measure is added to a regression that contains the other measures. We observe that $C$ keeps partial correlations above 0.54 after either control and raises the $R^{2}$ by 0.245 with the designed gaps and by 0.243 with the computed gaps, while novelty and unexpectedness raise it by at most 0.027. Therefore, $C$ passes the discriminant test, since it explains variation in reported curiosity that novelty and unexpectedness cannot, so it is a distinct construct.

\paragraph{(c) Testing Axiom 1: invariance to popularity and framing.}
Axiom \ref{ax:relativity} predicts that films matched on $(g,f,\tau)$ but differing in popularity and framing generate the same level of curiosity. 
We find that a matched pair of films from our candidate pool evokes the same curiosity (mean $1.61$ vs.\ $1.60$, $p = 0.82$), and we confirm with two one-sided tests that the difference lies within $\pm 0.5$ stars ($p < 0.001$).

\paragraph{(d) Testing Axiom 2: the interior peak.}
Axiom \ref{ax:perceivability} predicts that curiosity vanishes at zero context and at very large gaps, and therefore peaks at an intermediate gap. We find that participants report almost no curiosity for the zero-context movies at any gap level ($1.41$ stars on the seven-star scale), and little curiosity for the movie most similar to one they have already seen ($1.48$ stars). Between these two ends, curiosity first rises and then falls with the gap in both context scenarios ($p < 0.001$), with its peak at a gap of $2.80$ under low context. Meanwhile, perceived novelty keeps rising with the gap (mean Spearman correlation of 0.86), so a measure based on novelty cannot fit this pattern.

\paragraph{(e) Testing Axiom 3: peak location shifts with context.}
Axiom \ref{ax:optimum} says that the curiosity peak is located at a smaller gap when the consumer knows more about the item. We locate the peak separately in the low- and high-context scenarios. It sits at a gap of $2.80$ at low context and $1.87$ at high context, a shift of $0.93$ in the predicted direction, as reported in Table \ref{tab:peakshift}. The shift grows with the reference gap, from $-0.38$ in the low-$\tau$ tercile to $-0.87$ and $-1.82$ in the mid- and high-$\tau$ terciles, as the canonical form predicts, since the peak $\tau(1-f)$ shifts with the surrounding context in proportion to $\tau$. This peak shift is the observation that clearly separates curiosity from existing proxy metrics, none of which has a peak that shifts with context.

\paragraph{(f) Testing Axiom 4: peak location tracks the reference gap.}
Axiom \ref{ax:tracking} says that the peak sits at a larger gap when the reference gap $\tau$ is larger. We sort participants into terciles of the reference-gap proxy and locate each tercile's low-context peak. The peak rises across the terciles ($1.72$, $2.70$, and $3.73$, with a Spearman correlation of 0.71 and $p < 0.001$), as the axiom predicts. 

Finally, we test Axiom \ref{ax:recency}, which defines how $\tau$ changes over time, in the offline ablation of Section \ref{sec:ablation} and the online experiment of Section \ref{sec:fieldhet}.

\begin{table}[tbp]
\TABLE
{Construct and Discriminant Validity by Specification\label{tab:study1}}
{\begin{tabular}{@{}lcccc@{}}
\hline\up
Specification & \shortstack{Correlation\\with reported} & \shortstack{Partial,\\net novelty} & \shortstack{Partial,\\net unexp.} & \shortstack{Incremental\\$\Delta R^{2}$} \\ \hline\up
Canonical, designed $g$ & 0.636 & 0.562 & 0.557 & 0.245 \\
\quad Canonical, model-computed $g$      & 0.598 & 0.553 & 0.551 & 0.243 \\[2pt]
Perceived novelty                        & $-0.364$ & --- & $-0.154$ & 0.020 \\
Perceived unexpectedness                 & $-0.372$ & $-0.177$ & --- & 0.027 \down\\ \hline
\end{tabular}}
{}
\end{table}

\begin{table}[tbp]
\TABLE
{Curiosity-Maximizing Gap $\gs$ by Context and Reference Gap\label{tab:peakshift}}
{\begin{tabular}{@{}lccc@{}}
\hline\up
& Low $f$ & High $f$ & \shortstack{Difference\\(pred.\ $<0$)} \\ \hline\up
$\gs$ (all $\tau$)          & 2.80 & 1.87 & $-0.93$ \\
\quad Low-$\tau$ tercile    & 1.72 & 1.34 & $-0.38$ \\
\quad Mid-$\tau$ tercile    & 2.70 & 1.83 & $-0.87$ \\
\quad High-$\tau$ tercile   & 3.73 & 1.91 & $-1.82$ \down\\ \hline
\end{tabular}}
{}
\end{table}

\section{Offline Experiments}\label{sec:offline}

\subsection{Datasets and Baselines}\label{sec:datasets}
To validate the proposed recommendation framework, we conducted offline experiments on three real-world datasets with different recommendation scenarios (local business reviews, movies, and short-video streaming), and we show descriptive statistics in Table \ref{tab:datasets}. The first two are the public Yelp and MovieLens datasets, and the third is a proprietary short-video dataset from Alibaba-Youku. Each record contains consumer/item identifiers and ratings, which we transform to binary labels following common practice \citep{ricci2015}. The three datasets also differ in sparsity levels.
\begin{table}[tbp]
\TABLE
{Descriptive Statistics of the Datasets\label{tab:datasets}}
{\begin{tabular}{@{}lccc@{}}
\hline\up
Dataset & Yelp & MovieLens & Alibaba-Youku \\ \hline\up
\#Consumers & 41{,}607 & 129{,}757 & 6{,}110 \\
\#Items  & 26{,}487 & 11{,}508 & 4{,}404 \\
\#Records   & 1{,}431{,}100 & 19{,}625{,}886 & 247{,}623 \\
Sparsity    & 99.87\% & 98.69\% & 99.08\% \down\\ \hline
\end{tabular}}
{}
\end{table}

We compare our method against the following groups of baselines, listed in the order of Table \ref{tab:offline}:
\begin{itemize}
\item \textbf{Relevance-only recommender systems:} \emph{DIN} \citep{zhou2018}, which learns consumer interest from historical behavior through a local activation unit, and \emph{MF} \citep{koren2009}, which factorizes the consumer--item matrix into latent factors.
\item \textbf{Deviation-based proxy metrics:} \emph{Unexpectedness} \citep{adamopoulos2014}, which scores an item by its distance from the consumer's expected set with a preferred distance for each consumer, and \emph{Variety-seeking} \citep{li2024}, which adjusts the degree of unexpectedness by a consumer-level variety-seeking trait.
\item \textbf{Curiosity proxy:} \emph{PCC} \citep{fu2024}, which adopts the Wundt-curve model introduced in Section \ref{sec:notdeviation} to recommend items near the peak of a per-consumer curiosity curve over surprise.
\item \textbf{Diversity re-ranking methods:} \emph{MMR} \citep{carbonell1998}, which selects the next item by trading off its relevance against its similarity to the items already selected, and \emph{DPP} \citep{chen2018dpp}, which selects a list of relevant and mutually dissimilar items through a determinantal point process.
\end{itemize}

To ensure a fair comparison, each proxy metric is added as the second objective in the identical scalarization, and every model uses the same AutoEncoder item vectors of Section \ref{sec:gap}, so representation quality is held constant across the comparison. MMR and DPP re-rank the candidates of the base model with their own trade-off parameter between relevance and diversity, which we tune in the same way as $\alpha$. We tune the parameters and mixing weight of each baseline on the validation partition under the same tuning budget.

\subsection{Metrics and Hyperparameters}\label{sec:metrics}

Regarding the recommendation quality, we report \textbf{Hit Rate@10} (the share of events whose true item appears among the top ten recommendations) and \textbf{NDCG@10} (normalized discounted cumulative gain at ten, its position-discounted counterpart) as accuracy, and two measures beyond accuracy, including \textbf{Diversity} \citep{gunawardana2015}, the mean pairwise Euclidean distance between the item vectors of the top-ten items, and \textbf{Novelty} \citep{adamopoulos2014}, the share of the top-ten items that the consumer does not expect, based on her/his consumption history. We report results averaged over four time-stratified splits and five independent runs.

Hyperparameters are selected by grid search \citep{bergstra2012} on a validation partition of the training window, with a matched budget across methods. In particular, we determine by validation Hit Rate@10 the kernel bandwidth $\rho$ in $f$ (a multiple of the median nearest-neighbor distance), the decay $\lambda$ in $\tau$, and the width $\sigma$ in the canonical form. The mixing weight $\alpha$ is likewise selected per method, so the comparison is at each method's best weight. The selected values are stable across splits in our empirical analyses.

\subsection{Experiment Results}\label{sec:offlineresults}

Table \ref{tab:offline} reports the comparison. The pattern is consistent across all three datasets and both models, as every method improves Hit Rate@10, NDCG, and diversity over the base model, which is DIN in Panel (a) and MF in Panel (b).
Further, the calibrated curiosity objective achieves the largest improvement on every dataset and every metric, as the bold last row of each panel shows. The deviation-based proxy metrics, unexpectedness and variety-seeking, raise Hit Rate@10 over the base model but trail the curiosity objective on every dataset. PCC, the curiosity-proxy competitor, captures part of the nonmonotone structure through its Wundt-curve response, but it also trails $C$ in recommendation accuracy on every dataset. It is worth noting that the diversity re-ranking methods, MMR and DPP, also improve over the base model, yet both trail the curiosity objective in Hit Rate@10 and in diversity on every dataset. The curiosity objective also achieves the highest diversity and novelty on every dataset, and its advantage carries over to the non-sequential MF model in Panel (b). Together, these results show that our curiosity-aware recommender system improves accuracy, diversity, and novelty \textit{simultaneously}.

\begin{table}[tbp]
\TABLE
{Offline Comparison: Curiosity Versus Deviation-Based Alternatives\label{tab:offline}}
{\footnotesize
\begin{tabular}{@{}l*{12}{c}@{}}
\hline
& \multicolumn{4}{c}{\up Yelp} & \multicolumn{4}{c}{MovieLens} & \multicolumn{4}{c}{Alibaba-Youku} \\
\cline{2-5}\cline{6-9}\cline{10-13}
& HR@10 & NDCG & Div. & Nov. & HR@10 & NDCG & Div. & Nov. & HR@10 & NDCG & Div. & Nov. \up\\ \hline
\multicolumn{13}{@{}l}{\up\emph{Panel (a): DIN model}} \\
DIN (base)             & 0.409 & 0.213 & 0.200 & 0.929 & 0.422 & 0.221 & 0.260 & 0.701 & 0.113 & 0.057 & 0.040 & 0.831\\
\quad + Unexpectedness & 0.471 & 0.231 & 0.450 & 0.931 & 0.477 & 0.275 & 0.352 & 0.698 & 0.151 & 0.076 & 0.070 & 0.835 \\
\quad + Variety-seeking & 0.491 & 0.241 & 0.430 & 0.931 & 0.472 & 0.280 & 0.346 & 0.703 & 0.163 & 0.087 & 0.070 & 0.838 \\
\quad + PCC            & 0.428 & 0.219 & 0.410 & 0.923 & 0.461 & 0.262 & 0.340 & 0.704 & 0.159 & 0.078 & 0.070 & 0.829 \\
\quad + MMR            & 0.468 & 0.235 & 0.441 & 0.921 & 0.468 & 0.273 & 0.346 & 0.696 & 0.148 & 0.073 & 0.050 & 0.821 \\
\quad + DPP            & 0.473 & 0.228 & 0.459 & 0.921 & 0.465 & 0.271 & 0.358 & 0.698 & 0.155 & 0.085 & 0.070 & 0.825 \\
\quad \textbf{+ Curiosity $C$} & \textbf{0.512***} & \textbf{0.276***} & \textbf{0.480***} & \textbf{0.934***} & \textbf{0.495***} & \textbf{0.301***} & \textbf{0.378***} & \textbf{0.706***} & \textbf{0.216***} & \textbf{0.114***} & \textbf{0.090***} & \textbf{0.842***} \\[2pt]
\multicolumn{13}{@{}l}{\emph{Panel (b): MF model}} \\
MF (base)              & 0.250 & 0.127 & 0.180 & 0.898 & 0.496 & 0.261 & 0.242 & 0.671 & 0.111 & 0.056 & 0.030 & 0.819 \\
\quad + Unexpectedness & 0.402 & 0.198 & 0.390 & 0.901 & 0.520 & 0.283 & 0.318 & 0.673 & 0.144 & 0.073 & 0.060 & 0.825 \\
\quad + Variety-seeking & 0.431 & 0.211 & 0.370 & 0.902 & 0.516 & 0.279 & 0.312 & 0.673 & 0.152 & 0.077 & 0.060 & 0.827 \\
\quad + PCC            & 0.359 & 0.176 & 0.350 & 0.890 & 0.509 & 0.272 & 0.306 & 0.665 & 0.149 & 0.075 & 0.050 & 0.817 \\
\quad + MMR            & 0.411 & 0.203 & 0.388 & 0.892 & 0.513 & 0.278 & 0.312 & 0.667 & 0.140 & 0.076 & 0.060 & 0.808 \\
\quad + DPP            & 0.417 & 0.205 & 0.395 & 0.893 & 0.516 & 0.281 & 0.318 & 0.667 & 0.143 & 0.077 & 0.060 & 0.811 \\
\quad \textbf{+ Curiosity $C$} & \textbf{0.500***} & \textbf{0.263***} & \textbf{0.420***} & \textbf{0.904***} & \textbf{0.534***} & \textbf{0.297***} & \textbf{0.336***} & \textbf{0.675***} & \textbf{0.169***} & \textbf{0.086***} & \textbf{0.080***} & \textbf{0.830***} \down\\ \hline
\end{tabular}}
{Note: $^{***}p<0.01$ in a paired $t$-test against the best baseline.}
\end{table}

\subsection{Ablation of the Curiosity Components}\label{sec:ablation}

To examine whether the offline gains come from the components that the axioms require, we rerun the comparison on the DIN model with the following four ablation models, each of which holds part of the curiosity measure fixed while keeping the canonical form:
\begin{itemize}
\item \textbf{Ablation Model 1:} we replace the surrounding context $f_{uit}$ with its sample median $\bar{f}$, so that the peak no longer depends on how much the consumer knows about the item, which removes the mechanism of Axioms \ref{ax:perceivability} and \ref{ax:optimum}.
\item \textbf{Ablation Model 2:} we set the decay to $\lambda = 1$, so that the reference gap becomes a lifetime average of encountered gaps without recency, which removes the mechanism of Axiom \ref{ax:recency}.
\item \textbf{Ablation Model 3:} we replace the reference gap $\tau_{ut}$ with its sample median $\bar{\tau}$, so that the peak no longer shifts with the consumer's recent consumption, which removes the mechanism of Axioms \ref{ax:tracking} and \ref{ax:recency}.
\item \textbf{Ablation Model 4:} we hold both $f$ and $\tau$ at their sample medians, which leaves a measure that depends on the information gap alone and places its peak at the same gap for every consumer.
\end{itemize}

As reported in Table \ref{tab:ablation}, we find that each component of the curiosity measure contributes to the offline gains, as every ablation model lowers Hit Rate@10, NDCG, diversity, and novelty on all three datasets. Ablation Model 2, which removes recency from the reference gap, leads to the smallest loss, with Hit Rate@10 on Yelp falling from 0.512 to 0.489, while Ablation Models 1 and 3 lead to larger losses (0.483 and 0.481, respectively). Ablation Model 4, which depends on the information gap alone, has the lowest Hit Rate@10 and NDCG on every dataset, and on Yelp and MovieLens, its Hit Rate@10 even falls below that of the best deviation-based proxy metric in Table \ref{tab:offline}. Therefore, the surrounding context and the adaptive reference gap account for a large portion of the advantage of the curiosity objective over the best baseline across all three datasets.

\begin{table}[tbp]
\TABLE
{Ablation of the Curiosity Components (DIN Model)\label{tab:ablation}}
{\footnotesize
\begin{tabular}{@{}l*{12}{c}@{}}
\hline
& \multicolumn{4}{c}{\up Yelp} & \multicolumn{4}{c}{MovieLens} & \multicolumn{4}{c}{Alibaba-Youku} \\
\cline{2-5}\cline{6-9}\cline{10-13}
& HR@10 & NDCG & Div. & Nov. & HR@10 & NDCG & Div. & Nov. & HR@10 & NDCG & Div. & Nov. \up\\ \hline\up
\textbf{Full curiosity measure $C$} & \textbf{0.512***} & \textbf{0.276***} & \textbf{0.480***} & \textbf{0.934***} & \textbf{0.495***} & \textbf{0.301***} & \textbf{0.378***} & \textbf{0.706***} & \textbf{0.216***} & \textbf{0.114***} & \textbf{0.090***} & \textbf{0.842***} \\
\quad Ablation Model 1 & 0.483 & 0.262 & 0.463 & 0.921 & 0.465 & 0.281 & 0.342 & 0.685 & 0.200 & 0.107 & 0.070 & 0.829 \\
\quad Ablation Model 2 & 0.489 & 0.264 & 0.469 & 0.925 & 0.470 & 0.285 & 0.348 & 0.688 & 0.203 & 0.109 & 0.070 & 0.831 \\
\quad Ablation Model 3 & 0.481 & 0.258 & 0.451 & 0.917 & 0.454 & 0.279 & 0.336 & 0.681 & 0.199 & 0.104 & 0.050 & 0.829 \\
\quad Ablation Model 4 & 0.473 & 0.253 & 0.447 & 0.915 & 0.451 & 0.270 & 0.326 & 0.683 & 0.193 & 0.101 & 0.050 & 0.827 \down\\ \hline
\end{tabular}}
{Note: $^{***}p<0.01$ in a paired $t$-test against the best ablation model.}
\end{table}

\subsection{The Shape of the Curiosity Response}\label{sec:shape}

\begin{figure}[tbp]
\FIGURE
{\includegraphics[width=0.62\linewidth]{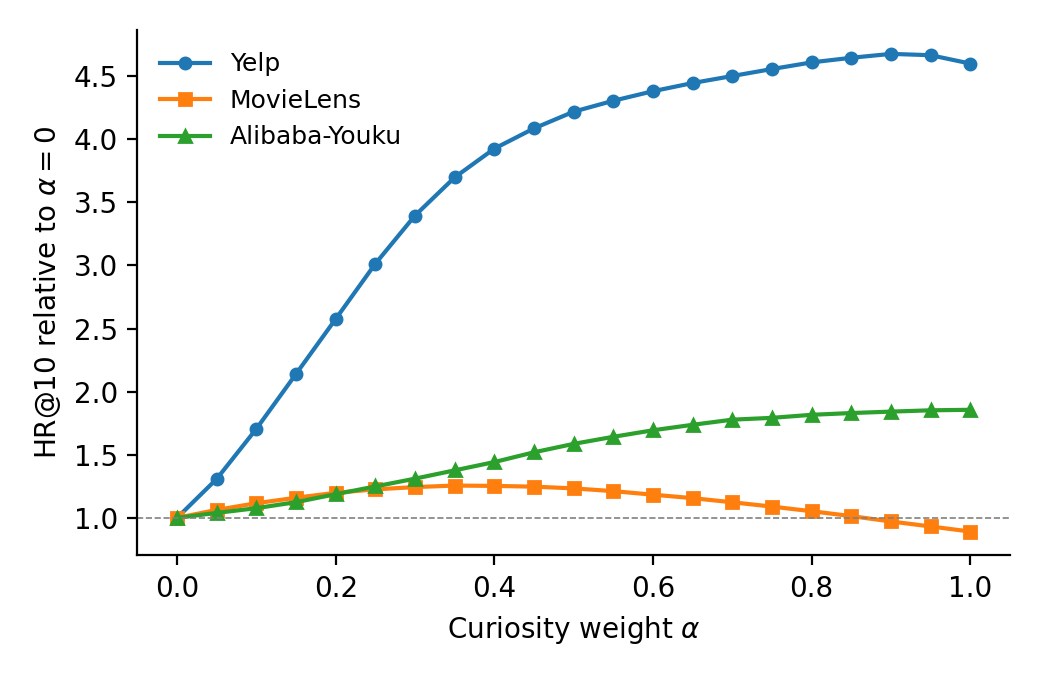}}
{Hit Rate@10 as a Function of the Curiosity Weight $\alpha$ (relative to
$\alpha=0$)\label{fig:shape}}
{}
\end{figure}

To examine how much weight the recommender system can give to curiosity, we vary the curiosity weight $\alpha$ in \eqref{eq:utility} from 0, where the ranking relies on relevance alone, to 1, where it relies on curiosity alone. Figure \ref{fig:shape} plots the Hit Rate@10 of the DIN model on each dataset relative to its value at $\alpha = 0$. We find that Hit Rate@10 on MovieLens rises to a peak near $\alpha = 0.4$ and falls as $\alpha$ approaches 1. On the sparser Yelp and Alibaba-Youku data, Hit Rate@10 peaks closer to $\alpha = 1$. Therefore, the best ranking on MovieLens combines relevance and curiosity, as its Hit Rate@10 at the peak exceeds that at both $\alpha = 0$ and $\alpha = 1$. Meanwhile, on Yelp and Alibaba-Youku, which have about 34 and 40 consumption records per consumer against 151 on MovieLens, a ranking that relies heavily on curiosity performs best. In practice, the curiosity weight thus needs to be carefully calibrated for each platform, as we do on the validation partition of Section \ref{sec:metrics}.

\subsection{Robustness Check}\label{sec:robustness}

The curiosity measure involves several implementation choices that the axioms leave open, and we examine whether our results depend on them. Since we have examined the item representation and the distance used to compute the information gap in Section \ref{sec:labresults}, the remaining choices are the computation of the surrounding context in Section \ref{sec:context}, the weights of the reference gap in Section \ref{sec:refgap}, and the scalarization of the utility function in Section \ref{sec:recsys}.

For these three choices, we rerun the offline experiment on the DIN model with the following three robustness checks, each of which replaces one choice of the primary specification with its alternative in Table \ref{tab:components}:
\begin{itemize}
\item \textbf{Robustness Check 1:} we compute the surrounding context $f$ with the $k$-nearest-neighbor measure, in place of the Gaussian kernel.
\item \textbf{Robustness Check 2:} we compute the reference gap $\tau$ with sliding-window weights, which weight the $W$ most recently consumed items equally, in place of the exponential decay, where we select the window length $W$ on the validation partition.
\item \textbf{Robustness Check 3:} we combine relevance and curiosity with a multiplicative scalarization in place of the linear one in \eqref{eq:utility}.
\end{itemize}

As we observe in Table \ref{tab:offlinerobust}, on every dataset, the Hit Rate@10 of each robustness check lies within $0.003$ of that of the primary specification. Therefore, the offline gains of the curiosity objective are robust to these alternative implementations of $f$, $\tau$, and the scalarization. These choices also leave the qualitative predictions of Section \ref{sec:canonical} unchanged, since every such prediction follows from the axioms alone.

\begin{table}[tbp]
\TABLE
{Offline Robustness of the Curiosity Objective (HR@10, DIN Model)\label{tab:offlinerobust}}
{\begin{tabular}{@{}lccc@{}}
\hline\up
Specification & Yelp & MovieLens & Alibaba-Youku \\ \hline\up
Primary specification & 0.512 & 0.495 & 0.216 \\
\quad Robustness Check 1 & 0.509 & 0.495 & 0.218 \\
\quad Robustness Check 2 & 0.512 & 0.495 & 0.217 \\
\quad Robustness Check 3 & 0.513 & 0.496 & 0.214 \down\\ \hline
\end{tabular}}
{Note: Each row changes one design choice of the primary specification.}
\end{table}

\subsection{Scalability}\label{sec:scalability}

\begin{figure}[tbp]
\FIGURE
{\includegraphics[width=0.85\linewidth]{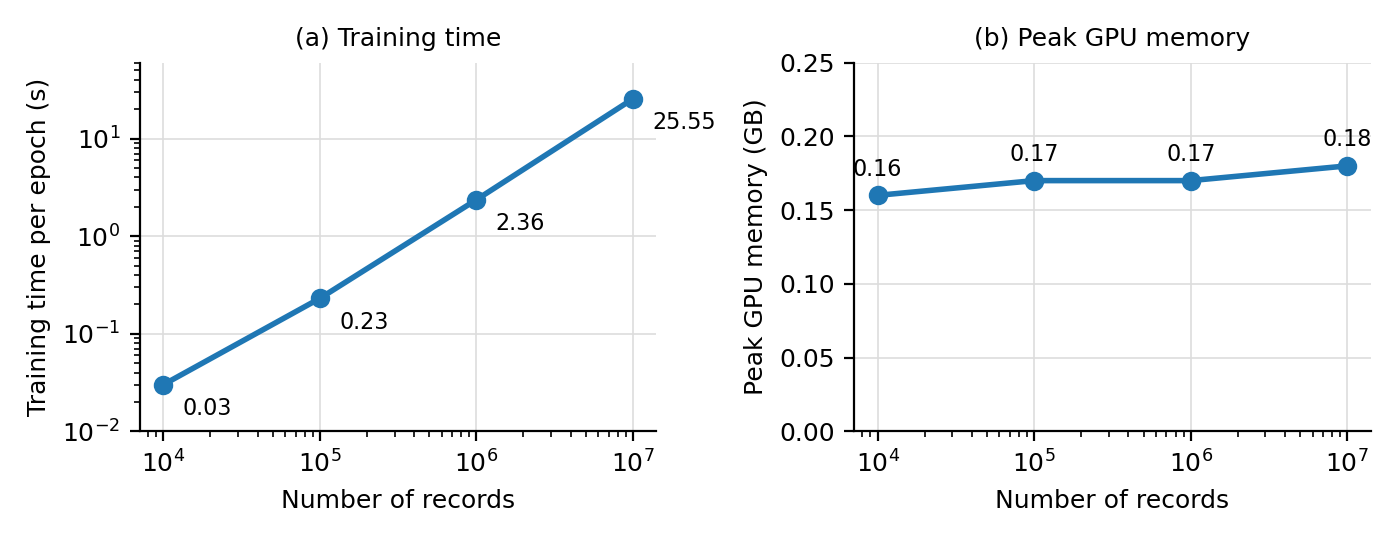}}
{Training Time per Epoch and Peak GPU Memory by Dataset Size\label{fig:scalability}}
{}
\end{figure}

To establish that the framework is deployable, we train the curiosity-aware model on randomly selected subsets of the data pooled across the three datasets spanning four orders of magnitude in size and record the training time per epoch on each. Figure \ref{fig:scalability} reports the results. Each tenfold increase in records raises the epoch time by a factor of 8 to 11, consistent with the $O(N)$ training complexity of the underlying deep learning techniques, while the peak GPU footprint stays between $0.16$ and $0.18$ GB because the model's parameters scale with the item vocabulary and not with the number of records. This linear scaling keeps the framework deployable at an industrial scale.

Incorporating the curiosity objective also adds little serving cost. The relevance model already maintains the item vectors and the sequential consumer state, and $g$, $f$, and $\tau$ are computed from these same quantities, where $\tau_{ut}$ is updated from its previous value and $f_{uit}$ requires only one kernel sum over the consumer's history for each candidate.

\section{Online Controlled Experiment}\label{sec:field}

To further validate our proposed recommendation framework in a real-world business setting, we conducted a large-scale online controlled experiment at one of the largest video-streaming platforms in Asia, with more than 500 million monthly active users and 800 million daily video views. Consumers receive video recommendations immediately on logging in, and may either click a recommended video or scroll down and refresh the page for a new set.

\subsection{Consumer Splitting and Identification Strategy}\label{sec:splitting}

The experiment drew 516{,}201 consumers from the platform's experimental pool and split them by binary hashing \citep{kohavi2009} into a control group, served by the production DIN system \citep{zhou2018}, and a treatment group, served by the curiosity-aware model of Section \ref{sec:recsys}, which augments the same production relevance score with the curiosity utility.\footnote{Due to the latency constraint of the production system, we use a simplified computation of the curiosity measure in the treatment group, which we describe in Appendix \ref{app:online} and which still satisfies all five axioms.} The interface was identical across the two groups, so consumers were unaware of their assignment, and the experiment ran over one month in 2021, comprising 26{,}959{,}538 recommendation records. Table \ref{tab:balance} reports covariate balance across the two groups, where we detect no significant difference with the Wilcoxon rank-sum test on any of the consumer features of Table \ref{tab:features}, which confirms that the randomization produced balanced groups. In addition, the standardized mean difference of every feature is at most 0.051, well below the threshold of 0.1 that is commonly used to indicate good balance \citep{austin2009}.

\begin{table}[tbp]
\TABLE
{Consumer Feature-Level Summary Statistics for the Control and Treatment Groups\label{tab:balance}}
{\footnotesize
\begin{tabular}{@{}llccccccc@{}}
\hline
\multicolumn{2}{c}{\up Consumer Features} & Mean & 25th pctl. & Median & 75th pctl. & Std.\ dev. & $p$-value & \shortstack{Std.\ mean\\diff.} \\ \hline\up
\multirow{2}{*}{Gender} & Treatment & 0.005 & $-1.000$ & 1.000 & 1.000 & 0.990 & \multirow{2}{*}{0.7040} & \multirow{2}{*}{0.002} \\
                        & Control   & 0.003 & $-1.000$ & 1.000 & 1.000 & 0.990 & & \\
\multirow{2}{*}{Age} & Treatment & 41.119 & 30.000 & 40.000 & 50.000 & 21.100 & \multirow{2}{*}{0.1787} & \multirow{2}{*}{0.051} \\
                     & Control   & 40.076 & 30.000 & 40.000 & 50.000 & 20.070 & & \\
\multirow{2}{*}{Province} & Treatment & 8.262 & 0.000 & 6.000 & 13.000 & 8.930 & \multirow{2}{*}{0.3898} & \multirow{2}{*}{0.023} \\
                             & Control   & 8.054 & 0.000 & 6.000 & 13.000 & 8.850 & & \\
\multirow{2}{*}{City} & Treatment & 2.038 & 1.000 & 2.000 & 3.000 & 1.550 & \multirow{2}{*}{0.8823} & \multirow{2}{*}{0.006} \\
                         & Control   & 2.028 & 1.000 & 2.000 & 3.000 & 1.530 & & \\
\multirow{2}{*}{Operating System} & Treatment & 1.172 & 1.000 & 1.000 & 1.000 & 0.410 & \multirow{2}{*}{0.8809} & \multirow{2}{*}{0.002} \\
                                  & Control   & 1.173 & 1.000 & 1.000 & 1.000 & 0.400 & & \\
\multirow{2}{*}{Premium Status} & Treatment & 0.361 & 0.000 & 0.000 & 1.000 & 0.480 & \multirow{2}{*}{0.4492} & \multirow{2}{*}{0.008} \\
                            & Control   & 0.365 & 0.000 & 0.000 & 1.000 & 0.480 & & \\
\multirow{2}{*}{Activity Metric} & Treatment & 20.668 & 8.000 & 19.000 & 27.000 & 8.520 & \multirow{2}{*}{0.8045} & \multirow{2}{*}{0.031} \\
                                 & Control   & 20.404 & 8.000 & 19.000 & 27.000 & 8.450 & & \down\\ \hline
\end{tabular}}
{Note: $p$-values are from the Wilcoxon rank-sum test, and the standardized mean difference divides the absolute difference in means by the pooled standard deviation \citep{austin2009}.}
\end{table}

\begin{table}[tbp]
\TABLE
{Descriptions of Consumer and Product Features Included in the Analysis Model\label{tab:features}}
{\footnotesize
\begin{tabular}{@{}llp{0.46\linewidth}l@{}}
\hline\up
Category & Variable & \multicolumn{1}{l}{Description} & Format \\ \hline\up
\multirow{7}{*}{\shortstack[l]{Consumer\\Features}}
  & Gender & Gender of the consumer & Categorical \\
  & Age & Age of the consumer & Numerical \\
  & Province & The province in which the consumer lives & Categorical \\
  & City & The city in which the consumer lives & Categorical \\
  & Operating System & The operating system on the consumer's device when logging into the platform & Categorical \\
  & Premium Status & Subscription to the premium service or not & Categorical \\
  & Activity Metric & A set of confidential features developed by the company to describe the level of consumer activities & Numerical \\[2pt] \hline
\multirow{5}{*}{\shortstack[l]{Video\\Features}}
  & Genre & The genre of the video & Categorical \\
  & View Count & Total number of views over the past month & Numerical \\
  & Comment Count & Total comment counts over the past month & Numerical \\
  & Release Days & The number of days since the video was released & Numerical \\
  & Content Metric & A set of confidential features developed by the company to describe video content & Numerical \down\\ \hline
\end{tabular}}
{}
\end{table}

\subsection{Average Treatment Effect}\label{sec:ate}

We measure three business outcomes, namely click-through rate ($CTR_{uit}$), which indicates whether the consumer clicked, video view ($VV_{uit}$), which indicates whether she/he finished the video, and time spent ($TS_{uit}$), the minutes watched. We estimate the treatment effect by ordinary least squares (OLS):
\begin{equation}\label{eq:ols}
\begin{aligned}
Metric_{uit} = {}& \beta_{0} + \beta_{1} Treatment_{u}\\
& + \overrightarrow{\beta_{2}}\overrightarrow{X_{i}}
+ \overrightarrow{\beta_{3}}\overrightarrow{Y_{u}} + D_{t} + \varepsilon_{uit},
\end{aligned}
\end{equation}
where $Metric_{uit} \in \{CTR_{uit}, VV_{uit}, TS_{uit}\}$, $Treatment_{u}$ is the assignment dummy, $\overrightarrow{X_{i}}$ and $\overrightarrow{Y_{u}}$ are the video and consumer features of Table \ref{tab:features}, $D_{t}$ are time fixed effects for date and hour, and $\varepsilon_{uit}$ is the error term. The coefficient $\beta_{1}$ measures the average treatment effect, and $\overrightarrow{\beta_{2}}$ and $\overrightarrow{\beta_{3}}$ are the coefficients of the video and consumer features.

\begin{table}[tbp]
\TABLE
{Average Treatment Effect of Curiosity-Aware Recommendations\label{tab:ate}}
{\begin{tabular}{@{}lccc@{}}
\hline\up
& CTR & VV & TS \\ \hline\up
Treatment           & 0.0275*** & 0.0197*** & 0.7073*** \\
                    & (0.001)   & (0.001)   & (0.0937)  \\
Consumer Features   & Yes & Yes & Yes \\
Video Features      & Yes & Yes & Yes \\
Time Fixed Effect   & Yes & Yes & Yes \\
R-Squared           & 0.0045 & 0.0050 & 0.0045 \\
Number of Consumers & 516{,}201 & 516{,}201 & 516{,}201 \\
Number of Records   & 26{,}959{,}538 & 26{,}959{,}538 & 26{,}959{,}538 \down\\ \hline
\end{tabular}}
{Note: Robust standard errors in parentheses. $^{***}p<0.01$.}
\end{table}

Table \ref{tab:ate} reports the estimates. The curiosity-aware system raises all three outcomes over the production baseline, as click-through rises by 2.75 percentage points, video views by 1.97 percentage points, and time spent by 0.71 minutes per recommendation (all $p < 0.01$). The magnitudes are economically meaningful against a mature production system. Given an average video length of about three minutes, the time-spent gain of 0.71 minutes per recommendation equals about one quarter of the length of an average video. We further conduct an illustrative calculation to translate these gains into the platform's total viewing activity. In our experiment, each consumer received 52 recommendations per month on average (26{,}959{,}538 records for 516{,}201 consumers). At the scale of all 500 million monthly active users, and if these consumers responded in the same way as those in our experiment, the time-spent gain would add 18.5 billion viewing minutes, or 308 million viewing hours, per month, and the video-view gain would add 514 million completed video views per month. To translate these gains into revenue, we use the public annual report of the platform's parent company, which reports US\$4.8 billion and US\$5.1 billion of revenue in fiscal years 2021 and 2022 for the business segment that includes the platform. Dividing this revenue by the platform's 292 billion annual video views (800 million per day) gives US\$0.016--0.017 of revenue per video view. If revenue grows in proportion to video views, the additional 6.2 billion completed video views per year would bring US\$101--108 million of additional revenue per year, or 2.1\% of the segment's annual revenue. These gains come from a single objective added on top of the existing relevance score, with no change to the production relevance model, which shows that the framework can be deployed in practice. Following the experiment, the platform deployed our curiosity-aware recommender system in production, where it remains in operation.

\subsection{Heterogeneous Treatment Effects}\label{sec:fieldhet}

To examine whether the treatment effect concentrates among the consumers in whom our theory predicts curiosity to be evoked most, we estimate how the effect varies with consumer attributes and with the state variables of the curiosity measure. Regarding consumer attributes, we consider age and premium status. Younger consumers are more exploratory \citep{raju1980}, so the treatment effect should decline with age. In addition, premium status grants access to the platform's larger exclusive library, which offers a wider set of novel content to be curious about. Regarding the state variables, we consider history depth, which is the number of previously watched videos, and the recency of the consumer's activity. Since Axiom~\ref{ax:perceivability} states that a gap is perceivable only where surrounding context exists, the effect should grow with history depth, and since Axiom~\ref{ax:recency} defines the reference gap $\tau$ as a recency-weighted average of encountered gaps, the effect should be larger for the more recently active consumers. Specifically, we estimate
\begin{equation}\label{eq:fieldhet}
\begin{aligned}
Metric_{uit} = {}& \beta_{0} + \beta_{1} Treatment_{u}\\
& + \beta_{2}\,(Treatment_{u} \times M_{u}) + \beta_{3} M_{u}\\
& + \overrightarrow{\beta_{4}}\overrightarrow{X_{i}}
  + \overrightarrow{\beta_{5}}\overrightarrow{Y_{u}} + D_{t} + \varepsilon_{uit},
\end{aligned}
\end{equation}
where $M_{u}$ is, in turn, the consumer's age group, premium status, history depth, and recency, each entered in a separate specification, $\beta_{2}$ measures how the treatment effect varies with $M_{u}$, $\beta_{3}$ is the main effect of $M_{u}$, and the remaining terms are as in Equation \eqref{eq:ols}.

Table \ref{tab:fieldhet} reports the interaction estimates, where we observe that the treatment effect falls monotonically with age. Compared with consumers who do not report their age, consumers under 30 gain 11.46 more percentage points in click-through and those between 30 and 60 gain 5.78 more points, while the interaction for consumers over 60 is significantly negative ($-13.84$ points), in line with the lower exploratory tendency of older consumers \citep{raju1980}. Premium consumers gain more than non-premium consumers on all three outcomes, with 1.24 more percentage points in click-through, 2.50 more points in video views, and 1.24 more minutes watched. Regarding the state variables, we find that both interactions are positive and significant, as a consumer one standard deviation deeper in history gains 0.87 more percentage points in click-through, and a consumer one standard deviation more recently active gains 1.19 more points. Therefore, the treatment effect is larger for younger and premium consumers and for consumers with deeper and more recent consumption histories, as our theory predicts.

In addition, we estimate the same specification with gender, city tier, and operating system as moderators, and Table \ref{tab:fielddemo} reports the interaction estimates. We find that the gender difference is insignificant on all three outcomes, that the treatment effect is significantly lower for consumers in the largest cities than for consumers with no city specified, and that consumers on iOS and Android gain substantially more than those on other systems.

\begin{table}[tbp]
\TABLE
{Heterogeneous Treatment Effects by Consumer Disposition\label{tab:fieldhet}}
{\begin{tabular}{@{}lccc@{}}
\hline\up
& CTR & VV & TS \\ \hline
\multicolumn{4}{@{}l}{\up\emph{Panel (a): Age group}} \\
Treatment $\times$ Under 30       & 0.1146*** & 0.1149*** & 3.0990*** \\
                                  & (0.002)   & (0.002)   & (0.2125)  \\
Treatment $\times$ 30--60         & 0.0578*** & 0.0473*** & 1.3854*** \\
                                  & (0.002)   & (0.003)   & (0.3027)  \\
Treatment $\times$ Above 60       & $-0.1384$*** & $-0.1332$*** & $-1.3855$*** \\
                                  & (0.003)   & (0.003)   & (0.3322)  \\
\multicolumn{4}{@{}l}{\up\emph{Panel (b): Premium status}} \\
Treatment $\times$ Premium        & 0.0124*** & 0.0250*** & 1.2375*** \\
                                  & (0.002)   & (0.002)   & (0.1937)  \\
\multicolumn{4}{@{}l}{\up\emph{Panel (c): Curiosity state (standardized, per SD)}} \\
Treatment $\times$ History Depth  & 0.0087*** & 0.0073*** & 0.2054*** \\
                                  & (0.002)   & (0.002)   & (0.061)   \\
Treatment $\times$ Recency        & 0.0119*** & 0.0109*** & 0.3038*** \\
                                  & (0.002)   & (0.003)   & (0.072)   \\ \hline\up
Main effects (moderators)         & Yes & Yes & Yes \\
Consumer \& Video Features        & Yes & Yes & Yes \\
Time Fixed Effect                 & Yes & Yes & Yes \\
Number of Records                 & 26{,}959{,}538 & 26{,}959{,}538 & 26{,}959{,}538 \down\\ \hline
\end{tabular}}
{Note: The omitted age group is consumers who do not report their age. Robust standard errors in parentheses. $^{*}p<0.1$; $^{**}p<0.05$; $^{***}p<0.01$.}
\end{table}

\begin{table}[tbp]
\TABLE
{Additional Heterogeneous Treatment Effects\label{tab:fielddemo}}
{\begin{tabular}{@{}lccc@{}}
\hline\up
& CTR & VV & TS \\ \hline
\multicolumn{4}{@{}l}{\up\emph{Panel (a): Gender}} \\
Treatment $\times$ Female         & $-0.0089$ & $-0.0091$ & $-0.3274$ \\
                                  & (0.006)   & (0.006)   & (0.5889)  \\
\multicolumn{4}{@{}l}{\up\emph{Panel (b): City tier}} \\
Treatment $\times$ Big City       & $-0.0915$*** & $-0.0940$*** & $-0.7890$*** \\
                                  & (0.002)   & (0.002)   & (0.2117)  \\
Treatment $\times$ Medium City    & $-0.0006$ & $-0.0005$ & 0.4785    \\
                                  & (0.004)   & (0.004)   & (0.3393)  \\
Treatment $\times$ Small City     & 0.0011    & $-0.0038$ & $-0.4886$ \\
                                  & (0.004)   & (0.004)   & (0.3593)  \\
\multicolumn{4}{@{}l}{\up\emph{Panel (c): Operating system}} \\
Treatment $\times$ Android        & 0.0248*** & 0.0226*** & 0.4866*** \\
                                  & (0.0009)  & (0.0009)  & (0.0540)  \\
Treatment $\times$ iOS            & 0.0760*** & 0.0758*** & 1.0361*** \\
                                  & (0.0009)  & (0.0009)  & (0.0570)  \down\\ \hline
\end{tabular}}
{Note: The omitted categories are records with no city specified in panel (b) and ``other'' operating systems in panel (c). Robust standard errors in parentheses. $^{*}p<0.1$; $^{**}p<0.05$; $^{***}p<0.01$.}
\end{table}

\subsection{Robustness Check}\label{sec:fieldrobust}

We test the robustness of our estimates under four additional settings, where our main empirical findings all stay the same:

(a) including different combinations of consumer features, video features, and time fixed effects.

(b) using the alternative discrete choice models of Logit and Probit \citep{hensher2005} to specify the binary outcomes $CTR_{uit}$ and $VV_{uit}$.

(c) excluding records of videos produced by the platform itself, to rule out that loyalty to the platform drives the clicks on these videos.

(d) excluding records from regions where the platform was launched recently, to rule out that the novelty of the platform itself drives the clicks there.

Table \ref{tab:fr-feat} reports check (a), and Table \ref{tab:fr-alt} reports checks (b) to (d). The treatment coefficient retains its sign, magnitude, and significance across all of them, so the average treatment effect of Table \ref{tab:ate} is not an artifact of the baseline specification.

\begin{table}[tbp]
\TABLE
{Average Treatment Effect on CTR under Different Feature and Fixed-Effect
Combinations\label{tab:fr-feat}}
{\begin{tabular}{@{}lccccc@{}}
\hline\up
& CTR & CTR & CTR & CTR & CTR \\ \hline\up
Treatment         & 0.0275*** & 0.0276*** & 0.0277*** & 0.0270*** & 0.0271*** \\
                  & (0.001)   & (0.001)   & (0.001)   & (0.001)   & (0.001)   \\
Consumer Features & Yes & No  & Yes & Yes & No  \\
Video Features    & Yes & Yes & No  & Yes & No  \\
Time Fixed Effect & Yes & Yes & Yes & No  & No  \\
R-Squared         & 0.0045 & 0.0019 & 0.0033 & 0.0039 & 0.0011 \\
Number of Consumers & 516{,}201 & 516{,}201 & 516{,}201 & 516{,}201 & 516{,}201 \\
Number of Records   & 26{,}959{,}538 & 26{,}959{,}538 & 26{,}959{,}538 & 26{,}959{,}538 & 26{,}959{,}538 \down\\ \hline
\end{tabular}}
{Note: Robust standard errors in parentheses. $^{*}p<0.1$; $^{**}p<0.05$; $^{***}p<0.01$.}
\end{table}

\begin{table}[tbp]
\TABLE
{Alternative Discrete-Choice Models and Subsamples\label{tab:fr-alt}}
{\begin{tabular}{@{}lccc@{}}
\hline\up
Specification & CTR & VV & TS \\ \hline
\multicolumn{4}{@{}l}{\up\emph{Discrete-choice models, check (b)}} \\
Logit  & 0.0274*** & 0.0196*** & --- \\
       & (0.001)   & (0.001)   &     \\
Probit & 0.0263*** & 0.0191*** & --- \\
       & (0.001)   & (0.001)   &     \\[2pt]
\multicolumn{4}{@{}l}{\emph{Subsamples, checks (c) and (d)}} \\
Excluding platform-produced videos  & 0.0271*** & 0.0197*** & 0.7068*** \\
                                    & (0.001)   & (0.001)   & (0.0944)  \\
Excluding recently launched regions & 0.0273*** & 0.0196*** & 0.7075*** \\
                                    & (0.001)   & (0.001)   & (0.0978) \down\\ \hline
\end{tabular}}
{Note: The discrete-choice models report average marginal effects for the binary outcomes. Robust standard errors in parentheses. $^{*}p<0.1$; $^{**}p<0.05$; $^{***}p<0.01$.}
\end{table}

\FloatBarrier
\section{Conclusions}\label{sec:conclusion}\label{sec:discussion}

Curiosity is an important driver of consumer decisions. In this paper, we formalized the information-gap theory of curiosity as a set of five axioms over the relation between an item's information and a consumer's knowledge state. We proved that these axioms, together with additional shape conditions, lead to a canonical form of the curiosity measure, which we validated through a survey-based study. Based on our measure, we subsequently constructed a curiosity-aware recommender system, and we demonstrated through extensive offline experiments that it significantly improves ranking performance on all three datasets and outperforms the unexpectedness, variety-seeking, Wundt-curve, and diversity re-ranking alternatives. Furthermore, through a large-scale online controlled experiment, we showed that the curiosity-aware recommender system significantly raises click-through rate, video views, and time spent over the latest production system, with the gains concentrated among the consumers in whom the theory predicts curiosity should be evoked most, by disposition and by knowledge state, and the platform has since deployed the system in production. Together, the three forms of validation establish that $C$ is a valid measure of curiosity and that incorporating it into recommendations raises consumer engagement.

Our findings lead to important managerial implications for the design of recommendation platforms. First, platforms can measure the curiosity that each candidate item evokes in each consumer directly from consumption records, using the item vectors and consumer states that their recommendation models already maintain, so the curiosity objective can be added to a deployed system without redesigning it and with little additional serving cost, as the deployment of our system at the video-streaming platform shows. Second, platforms should ask which item opens a gap that each consumer can perceive. Recommendations chosen this way outperform the deviation-based alternatives in the offline experiments and the latest production system in the online experiment. Third, platforms should personalize the degree of curiosity-based recommendation for each consumer, since the heterogeneous treatment effects show that it benefits younger, premium, and recently active consumers the most, and can backfire for consumers who prefer their familiar content, so the curiosity weight should be raised for the former group and reduced for the latter.

There are several directions in which future research can extend the current work. First, as our online experiment was conducted on one video-streaming platform, future research can examine curiosity-aware recommendations on other platforms and in other applications, including music and e-commerce. Second, the framework can be applied beyond recommendations, wherever a system chooses what information to place before someone who already knows something adjacent to it, such as search, education, and news feeds.

\bibliographystyle{informs2014}
\bibliography{references}

\newpage

\begin{APPENDICES}

\makeatletter
\@addtoreset{table}{section}
\@addtoreset{figure}{section}
\@addtoreset{equation}{section}
\@addtoreset{theorem}{section}
\@addtoreset{lemma}{section}
\@addtoreset{proposition}{section}
\@addtoreset{corollary}{section}
\@addtoreset{definition}{section}
\let\c@lemma\c@theorem
\let\c@proposition\c@theorem
\let\c@corollary\c@theorem
\let\c@definition\c@theorem
\makeatother
\renewcommand{\thetable}{\thesection.\arabic{table}}
\renewcommand{\thefigure}{\thesection.\arabic{figure}}
\renewcommand{\theequation}{\thesection.\arabic{equation}}
\renewcommand{\thetheorem}{\thesection.\arabic{theorem}}
\renewcommand{\thelemma}{\thesection.\arabic{lemma}}
\renewcommand{\theproposition}{\thesection.\arabic{proposition}}
\renewcommand{\thecorollary}{\thesection.\arabic{corollary}}
\renewcommand{\thedefinition}{\thesection.\arabic{definition}}

\section{Implicit Assumptions of the Axioms}\label{app:foundations}

Each axiom of Section \ref{sec:axioms}, together with the closure restriction of Section \ref{sec:prelim}, carries implicit assumptions that delimit the settings in which the framework applies. We record them here, with each axiom restated under its main-text number. The anchor quotations and the examples for each claim appear with the axioms in Section \ref{sec:axioms}, and the predictions of the axioms are tested in the survey-based study of Section \ref{sec:labstudy}, the offline experiments of Section \ref{sec:offline}, and the online controlled experiment of Section \ref{sec:field}.

\subsection{The Closure Restriction}

\noindent $C$ is defined only on $\Dom$, so an already-consumed item $i \in \Hut$ carries no curiosity value.

\noindent\textbf{Implicit assumptions.} (i) \emph{History observability}: the consumption history $\Hut$ is taken as known, and latent prior knowledge that the consumer brings from outside the platform is not represented. (ii) \emph{No forgetting}: a consumed item stays in the history within the modeling window. (iii) \emph{Binary consumption}: an item is either consumed or not.

\subsection{Axiom \ref{ax:relativity}: Gap Relativity}

\begin{repeataxiom}[Axiom \ref{ax:relativity} (Gap Relativity)]
There exists a continuous function $F : \Rplus \times [0,1] \times (0,\infty) \rightarrow \Rplus$ such that $C(u,i,t) = F\left( g_{uit},f_{uit},\tau_{ut} \right)$ for all $(u,i,t) \in \Dom$; i.e., curiosity depends on $(u,i,t)$ only through the gap-relative quantities $(g,f,\tau)$.
\end{repeataxiom}

\noindent\textbf{Implicit assumptions.} (i) \emph{Presentation-neutrality}: we model curiosity from the item's information, not from interface or presentation manipulations. (ii) \emph{Value-independence}: curiosity does not depend on the utility of the missing information.

\subsection{Axiom \ref{ax:perceivability}: Perceivability}

\begin{repeataxiom}[Axiom \ref{ax:perceivability} (Perceivability)]
$F(g,0,\tau) = 0$ for all $g,\tau$; and $\lim_{g \to \infty} F(g,f,\tau) = 0$ for each $f \in (0,1)$, $\tau > 0$.
\end{repeataxiom}

\noindent\textbf{Implicit assumptions.} \emph{Margins only}: the axiom constrains only the two extremes and asserts nothing about the interior shape, which the further conditions of Theorem \ref{thm:char} pin down.

\subsection{Axiom \ref{ax:optimum}: Context-Dependent Optimum}

\begin{repeataxiom}[Axiom \ref{ax:optimum} (Context-Dependent Optimum)]
For all $\tau > 0$ and $0 < f < f' < 1$: $\gs(f',\tau) < \gs(f,\tau)$.
\end{repeataxiom}

\noindent\textbf{Implicit assumptions.} \emph{Fixed reference gap}: the claim holds $\tau$ constant, isolating the context effect from the adaptive reference gap of Axiom \ref{ax:tracking}.

\subsection{Axiom \ref{ax:tracking}: Reference-Gap Tracking}

\begin{repeataxiom}[Axiom \ref{ax:tracking} (Reference-Gap Tracking)]
For each $f \in (0,1)$ and $0 < \tau < \tau'$: $\gs(f,\tau) < \gs(f,\tau')$.
\end{repeataxiom}

\noindent\textbf{Implicit assumptions.} \emph{Reference gap as location}: $\tau$ sets the axis position of the optimum, distinct from $f$, and the two are assumed separately identifiable, with $f$ shifting the peak by context and $\tau$ by the consumer's current reference gap.

\subsection{Axiom \ref{ax:recency}: Recency Adaptivity}

\begin{repeataxiom}[Axiom \ref{ax:recency} (Recency Adaptivity)]
$\tau_{ut} = \sum_{(j,t') \in \Hut} \omega_{jt'}\, g_{ujt'}$ with $\omega_{jt'} \geq 0$, $\sum \omega_{jt'} = 1$, weights non-increasing in elapsed time and not constant.
\end{repeataxiom}

\noindent\textbf{Implicit assumptions.} (i) \emph{Averaging form}: the axiom commits to a convex combination of encountered gaps but leaves the decay family (e.g., exponential weights) unspecified. (ii) \emph{Consumed items only}: $\tau$ is computed from consumed items, so it tracks what the consumer has absorbed and not the items that the platform only displayed.

\section{Proofs of the Theoretical Results}\label{app:proofs}

This appendix provides the proofs of the theoretical results in Section \ref{sec:construct}. In Appendix \ref{app:wellposed}, we first show that the curiosity-maximizing gap is well defined and that Axioms \ref{ax:relativity}--\ref{ax:tracking} lead to a location-family representation of the curiosity function, which we use to prove Theorem \ref{thm:char} in Appendix \ref{app:charproof}. We then discuss the four shape conditions of the theorem in Appendix \ref{app:conditions}, characterize the alternatives that arise when we relax one condition at a time in Appendix \ref{app:altproof}, verify that the canonical instantiation satisfies the five axioms in Appendix \ref{app:verify}, and prove in Appendix \ref{app:invariance} that every curiosity function that satisfies the axioms produces the same qualitative predictions.

\subsection{Well-Posedness and the Location-Family Representation}\label{app:wellposed}

In Lemma \ref{lem:wellposed}, we first show that the curiosity-maximizing gap $\gs$ exists at every $(f,\tau)$, and in Proposition \ref{prop:repr}, we show that Axioms \ref{ax:relativity}--\ref{ax:tracking} represent the curiosity function as a peaked profile recentered at a location surface.

We define $\Gs(f,\tau) := \arg\max_{g \geq 0} F(g,f,\tau)$ and $\gs(f,\tau) := \max \Gs(f,\tau)$. Increasing and decreasing are meant in the strong sense in this appendix, so that, for example, $m$ is decreasing in $f$ if $f < f'$ implies $m(f',\tau) < m(f,\tau)$.

\begin{definition}[Peaked profile; location surface]\label{def:profile}
A \textbf{peaked profile} on $[-y_{0},\infty)$, for some $y_{0} \geq 0$, is a continuous $\varphi : [-y_{0},\infty) \rightarrow \Rplus$, not identically zero, reaching its maximum at $0$ with $0$ its largest maximizer, and vanishing at infinity. A \textbf{location surface} is a continuous $m : (0,1) \times (0,\infty) \rightarrow \Rplus$, decreasing in $f$ and increasing in $\tau$.
\end{definition}

\begin{lemma}\label{lem:wellposed}
For each $f \in (0,1)$, $\tau > 0$, the following are equivalent: (i) the profile $g \mapsto F(g,f,\tau)$ is not identically zero; (ii) $\Gs(f,\tau)$ is nonempty and bounded, in which case it is compact and $\gs$ exists, with $F(\gs,f,\tau) > 0$. Since Axioms \ref{ax:optimum}--\ref{ax:tracking} quantify over $\gs(f,\tau)$ at every $(f,\tau)$, the axiom system presupposes (ii) throughout.
\end{lemma}

\begin{proof}{Proof}
(i) $\Rightarrow$ (ii): let $M = \sup_{g \geq 0} F(g,f,\tau) > 0$. By Axiom \ref{ax:perceivability} there is $G$ with $F < M/2$ for $g > G$, so $\sup_{[0,G]} F = M$, reached by continuity on the compact $[0,G]$; $\Gs = \{ F = M\} \cap [0,G]$ is nonempty, closed, and bounded. (ii) $\Rightarrow$ (i): if $F(\cdot,f,\tau) \equiv 0$, every $g \geq 0$ is a maximizer, so $\Gs = [0,\infty)$ is unbounded, contradicting (ii).\Halmos
\end{proof}

\begin{proposition}[Location-family representation]\label{prop:repr}
A curiosity function satisfies Axioms \ref{ax:relativity}, \ref{ax:perceivability}, \ref{ax:optimum}, \ref{ax:tracking} and has a continuous curiosity-maximizing gap $\gs$ on $(0,1) \times (0,\infty)$ if and only if $F(g,f,\tau) = \varphi_{f,\tau}\left( g - m(f,\tau) \right)$ for $f \in (0,1)$ and $F(g,0,\tau) = 0$, where $m$ is a location surface and $\{\varphi_{f,\tau}\}$ a jointly continuous family of peaked profiles, each on $[-m(f,\tau),\infty)$, such that the resulting $F$ is continuous at $f = 0$ and extends continuously to $f = 1$. The representation is unique: $m = \gs$ and $\varphi_{f,\tau}(y) = F\left( y + m(f,\tau),f,\tau \right)$.
\end{proposition}

\begin{proof}{Proof}
\emph{Necessity.} By Lemma \ref{lem:wellposed}, $\gs$ exists on $(0,1) \times (0,\infty)$; set $m := \gs$ and $\varphi_{f,\tau}(y) := F\left( y + m(f,\tau),f,\tau \right)$ on $y \geq -m(f,\tau)$, so $F = \varphi_{f,\tau}(g - m)$ identically. $\varphi_{f,\tau}(0) = \max_{g} F > 0$; any maximizer $y'$ of $\varphi_{f,\tau}$ has $y' + m \in \Gs$, hence $y' \leq \max\Gs - m = 0$, so $0$ is the largest maximizer; $\varphi_{f,\tau}(y) \rightarrow 0$ as $y \rightarrow \infty$ by Axiom \ref{ax:perceivability}; continuity and non-negativity are inherited. Thus each $\varphi_{f,\tau}$ is a peaked profile. Axioms \ref{ax:optimum}--\ref{ax:tracking} state that $m = \gs$ is decreasing in $f$ and increasing in $\tau$, and $m$ is continuous by assumption. The $f = 0$ clause is Axiom \ref{ax:perceivability}'s near margin, and the continuity of $F$ at $f = 0$ and $f = 1$ is part of Axiom \ref{ax:relativity}.

\emph{Sufficiency.} Given a location surface $m$ and peaked profiles $\{\varphi_{f,\tau}\}$, define $F$ by the display and extend it continuously to $f = 1$. Then $F$ is continuous, satisfies Axiom \ref{ax:relativity} by construction and Axiom \ref{ax:perceivability} at both margins, and the largest maximizer of $F(\cdot,f,\tau)$ is $m(f,\tau) + 0$, so $\gs = m$ is continuous and Axioms \ref{ax:optimum}--\ref{ax:tracking} follow from the monotonicity of $m$.

\emph{Uniqueness.} Any representation with the stated properties forces $\widetilde{m} = \gs = m$, and then $\widetilde{\varphi}_{f,\tau}(y) = F(y + m,f,\tau) = \varphi_{f,\tau}(y)$.\Halmos
\end{proof}

\begin{corollary}[Interior optimum]\label{cor:interior}
For every $f \in (0,1)$, $\tau > 0$: $0 < \gs(f,\tau) < \infty$.
\end{corollary}

\begin{proof}{Proof}
Existence is Lemma \ref{lem:wellposed}; for $f' \in (f,1)$, Axiom \ref{ax:optimum} gives $\gs(f,\tau) > \gs(f',\tau) \geq 0$.\Halmos
\end{proof}

\subsection{Proof of the Characterization Theorem}\label{app:charproof}

We state Theorem \ref{thm:char} with the shape conditions (S1)--(S4) in full and prove it in Steps 0--9 below, where Step 0 shows that the gate is positive, Steps 1--4 determine the profile $\psi$, Steps 5--7 the peak location $m$, Step 8 the gate $A$, and Step 9 assembles the three.

\begin{repeattheorem}[Theorem \ref{thm:char} (Characterization)]
Let $C$ satisfy Axioms \ref{ax:relativity}--\ref{ax:recency} and let $F$ be the function of Axiom \ref{ax:relativity}. The \textit{gate} $A(f)$ is the highest curiosity level that an item with surrounding context $f$ can evoke. The \textit{peak location} $m(f,\tau)$ is the gap at which this highest level is reached, and it coincides with the curiosity-maximizing gap $\gs(f,\tau)$. The \textit{profile} $\psi$ is the common shape by which curiosity falls as the gap $g$ gets farther from the peak, normalized so that $\psi(0) = 1$ at the peak. Consider the following four conditions:

\textbf{(S1) Separation.} $F(g,f,\tau) = A(f)\,\psi\left( g - m(f,\tau) \right)$ on $f \in (0,1)$, where $\psi : \mathbb{R} \rightarrow \Rplus$ is continuous with $\psi(0) = 1$ and is independent of $(f,\tau)$, and $A$ is independent of $\tau$;

\textbf{(S2) Orthogonal decomposability.} $\psi\left( \sqrt{x^{2} + y^{2}} \right) = \psi(x)\,\psi(y)$ for all $x,y \in \mathbb{R}$; that is, the profile treats two orthogonal components of a deviation independently;

\textbf{(S3) Location structure.} $m(f,\gamma\tau) = \gamma\, m(f,\tau)$ for all $\gamma > 0$; the increment $\mu(f + \delta) - \mu(f)$ is independent of $f$ for all $f$ and $\delta$ with $f, f + \delta \in (0,1)$, where $\mu(f) := m(f,1)$; and $\mu(0^{+}) = 1$ and $\mu(1^{-}) = 0$ for the one-sided limits of $\mu$ at $0$ and $1$;

\textbf{(S4) Gate structure.} $A(f + \delta) - A(f)$ is independent of $f$ for all $f$ and $\delta$ with $f, f + \delta \in (0,1)$, and $A(1^{-}) = 1$.

Then $C$ satisfies conditions (S1)--(S4) if and only if there exists a unique $\sigma^{2} > 0$ such that
\[
F(g,f,\tau) \;=\; f \cdot \exp\left( - \frac{(g - \tau(1 - f))^{2}}{2\sigma^{2}} \right).
\]
\end{repeattheorem}

\begin{proof}{Proof}
We fix $\tau > 0$ where a single value suffices, and all steps are uniform in $\tau$.

\emph{Step 0 (positivity of the gate on $(0,1)$).} By Lemma \ref{lem:wellposed} the maximum of $F(\cdot,f,\tau)$ is positive for each $f \in (0,1)$, and by the definition of the peak location this maximum is reached at $m(f,\tau) = \gs(f,\tau)$. Evaluating (S1) at $g = m(f,\tau)$ gives $F\left( m(f,\tau),f,\tau \right) = A(f)\,\psi(0) = A(f)$. Hence $A(f) > 0$ on $(0,1)$.

\emph{Step 1 (evenness of $\psi$).} Set $y = 0$ in (S2): $\psi(|x|) = \psi\left( \sqrt{x^{2}} \right) = \psi(x)\,\psi(0) = \psi(x)$ for all $x$. Hence $\psi(x) = \psi(|x|)$: $\psi$ is even, and it suffices to determine $\psi$ on $\Rplus$.

\emph{Step 2 (positivity of $\psi$).} Taking $x = y = z/\sqrt{2}$ in (S2): $\psi(z) = \psi\left( z/\sqrt{2} \right)^{2}$. If $\psi(z_{0}) = 0$ for some $z_{0}$, then $\psi\left( z_{0}/\sqrt{2} \right) = 0$, and by induction $\psi\left( z_{0}/2^{k/2} \right) = 0$ for every $k$; but $z_{0}/2^{k/2} \rightarrow 0$ and $\psi$ is continuous with $\psi(0) = 1$, a contradiction. Hence $\psi > 0$ everywhere, and $\log\psi$ is well-defined.

\emph{Step 3 (Cauchy equation and the Gaussian shape).} Define $q(t) := \log\psi\left( \sqrt{t} \right)$ for $t \geq 0$. For $s,t \geq 0$, applying (S2) with $x = \sqrt{s}$, $y = \sqrt{t}$:
\begin{align*}
q(s + t) &= \log\psi(\sqrt{s + t})
= \log[\psi(\sqrt{s})\,\psi(\sqrt{t})]\\
&= q(s) + q(t).
\end{align*}
Thus $q$ is additive on $\Rplus$ and continuous, so $q(t) = ct$ with $c = q(1)$ (additivity gives $q(t) = n\, q(t/n)$ and $q(rt) = r\, q(t)$ for rational $r$; continuity extends to all reals). Hence $\psi(y) = e^{cy^{2}}$ for $y \geq 0$, and by evenness for all $y \in \mathbb{R}$.

\emph{Step 4 (sign of $c$).} By Axiom \ref{ax:perceivability}, $F(g,f,\tau) = A(f)\,\psi(g - m) \rightarrow 0$ as $g \rightarrow \infty$ with $A(f) > 0$ (Step 0), so $\psi(y) \rightarrow 0$ as $y \rightarrow \infty$. If $c = 0$ then $\psi \equiv 1$; if $c > 0$ then $\psi \rightarrow \infty$; both contradict the limit. Hence $c < 0$; write $c = -1/(2\sigma^{2})$ with $\sigma^{2} = -1/(2c) > 0$. The sign condition determines $\psi$ uniquely up to the single width parameter: $\psi(y) = e^{-y^{2}/2\sigma^{2}}$, whose unique maximizer is $0$, consistent with Step 0.

\emph{Step 5 (homogeneity reduces the location surface).} For any $\tau > 0$, (S3) gives $m(f,\tau) = m(f,\tau \cdot 1) = \tau\, m(f,1) = \tau\,\mu(f)$.

\emph{Step 6 (uniform increments force affinity of $\mu$).} Let $\Delta(\delta) := \mu(f + \delta) - \mu(f)$, well-defined independently of $f$ by (S3). For $\delta_{1},\delta_{2} > 0$ with $f, f + \delta_{1} + \delta_{2} \in (0,1)$:
\begin{align*}
\Delta(\delta_{1} + \delta_{2})
&= [\mu(f + \delta_{1} + \delta_{2}) - \mu(f + \delta_{1})]\\
&\quad + [\mu(f + \delta_{1}) - \mu(f)]\\
&= \Delta(\delta_{2}) + \Delta(\delta_{1}).
\end{align*}
So $\Delta$ is additive where defined. Since $\mu(f) = m(f,1) = \gs(f,1)$ is decreasing in $f$ by Axiom \ref{ax:optimum}, $\Delta$ is also monotone, and a monotone additive function is linear, so $\Delta(\delta) = \mu_{1}\delta$ for a constant $\mu_{1}$. Hence $\mu$ is affine, $\mu(f) = \mu_{0} + \mu_{1} f$.

\emph{Step 7 (endpoints pin the location).} $\mu(0^{+}) = \mu_{0} = 1$ and $\mu(1^{-}) = \mu_{0} + \mu_{1} = 0$ give $\mu_{1} = -1$, so $\mu(f) = 1 - f$ and $m(f,\tau) = \tau(1 - f)$, which is decreasing in $f$, increasing in $\tau$, continuous, and positive on $(0,1)$, as Proposition \ref{prop:repr} requires.

\emph{Step 8 (the gate is linear).} The additivity argument of Step 6 applied to (S4) shows that $\Delta_{A}(\delta) := A(f + \delta) - A(f)$ is additive. Since $A > 0$ by Step 0, $\Delta_{A}(\delta) > -A(f)$ is bounded below for fixed $f$, and an additive function bounded below on an interval is linear, so $A$ is affine on $(0,1)$, $A(f) = A_{0} + A_{1} f$. For the intercept, continuity of $F$ and Axiom \ref{ax:perceivability}'s near margin give
\begin{align*}
A_{0} = \lim_{f \rightarrow 0^{+}} A(f)
&= \lim_{f \rightarrow 0^{+}} F(m(f,\tau),f,\tau)\\
&= F(\tau,0,\tau) = 0,
\end{align*}
using $m(f,\tau) \rightarrow \tau$ (Step 7) and $F(\cdot,0,\tau) \equiv 0$. Then $A(1^{-}) = A_{1} = 1$, so $A(f) = f$, which is positive on $(0,1)$ and consistent with Step 0.

\emph{Step 9 (assembly, uniqueness, and consistency).} Substituting Steps 4, 7, and 8 into (S1) yields the canonical form \eqref{eq:canonical} of the main text. Each factor was determined uniquely: $\psi$ within $\{ e^{-y^{2}/2\sigma^{2}} : \sigma^{2} > 0\}$ by Steps 1--4, $m$ by Steps 5--7, and $A$ by Step 8; hence $F$ is unique given $\sigma^{2}$. Conversely, the display satisfies (S1)--(S4): for (S2), $e^{-(x^{2} + y^{2})/2\sigma^{2}} = e^{-x^{2}/2\sigma^{2}}e^{-y^{2}/2\sigma^{2}}$, so the conditions are mutually consistent, and it satisfies Axioms \ref{ax:relativity}--\ref{ax:recency}, as we verify in Appendix \ref{app:verify}.\Halmos
\end{proof}

\subsection{Discussion of the Shape Conditions}\label{app:conditions}

The five axioms alone do not determine a unique curiosity function. By Proposition \ref{prop:repr}, they restrict the curiosity function to a peaked profile recentered at a location surface, but they leave the shape of the profile, the location of the peak, and the height of the peak free, so that infinitely many functions satisfy them. A recommender system, however, needs one specific function to compute the curiosity of every candidate item, and this function should have as few free parameters as possible, so that we can estimate it reliably from consumption records. The shape conditions serve this purpose. (S1) separates the height, the location, and the shape of the curve, so that each of them can be determined on its own, while (S2) fixes the shape, (S3) fixes the location, and (S4) fixes the height, which leaves only the width $\sigma^{2}$ to estimate. Each of (S2)--(S4) is necessary for this result, since relaxing any one of them yields a different family of curiosity functions that still satisfy the five axioms, as we show in Corollary \ref{cor:alternatives}. At the same time, the shape conditions do not affect the qualitative predictions that we test in this paper, since these predictions follow from the axioms alone, as we show in Proposition \ref{prop:invariance}. Therefore, the shape conditions select the most parsimonious member of the family without changing what the theory predicts.

The four conditions describe how the three components combine, and each is a natural choice for a curiosity measure.

\paragraph{(S1) Separation.} The surrounding context and the reference gap affect curiosity only through the gate $A(f)$, which sets the height of the curve, and the peak location $m(f,\tau)$, which sets where the curve peaks, while the profile $\psi$ is the same for all values of $f$ and $\tau$. Therefore, the decline in curiosity when the information gap $g$ differs from the peak location depends only on the difference $g - m(f,\tau)$, and it is the same for every knowledge state. This condition is what allows us to determine the shape, the location, and the height of the curve separately in (S2)--(S4).

\paragraph{(S2) Orthogonal decomposability.} The profile combines two deviations $x$ and $y$ in the same way as the Euclidean distance combines two orthogonal components, so that the profile at a deviation of size $\sqrt{x^{2}+y^{2}}$ equals the product of the profiles at $x$ and $y$, $\psi\left(\sqrt{x^{2}+y^{2}}\right) = \psi(x)\,\psi(y)$. Therefore, the decline in curiosity due to one component of the deviation does not depend on the other component. For example, an item that differs from the peak in both its era and its style loses curiosity for each difference separately. Together with the continuity of $\psi$, this condition fixes the profile as the Gaussian $\psi(y) = e^{-y^{2}/2\sigma^{2}}$, as we show in Steps 1--4 of the proof in Appendix \ref{app:charproof}. Without it, the profile can be any peaked profile, including compact-support kernels such as the Epanechnikov and triangular kernels, as well as asymmetric profiles, as we show in Corollary \ref{cor:alternatives}(i).

\paragraph{(S3) Location structure.} This condition determines where the peak sits. The reference gap $\tau$ is measured in the same units as the gap itself, so if $\tau$ doubles, the curiosity-maximizing gap doubles. At near-zero surrounding context, the peak is located at the reference gap $\tau$, at full context it is located at zero, and between these two endpoints each equal gain in the surrounding context brings the peak closer by an equal amount. Together, these requirements give the peak location $m(f,\tau) = \tau(1 - f)$, as we show in Steps 5--7 of the proof. If equal gains in the surrounding context reduce the remaining distance by equal proportions, the peak location becomes $m(f,\tau) = \tau(1-f)^{\kappa}$ for some $\kappa > 0$, and the canonical form is the case $\kappa = 1$, as we show in Corollary \ref{cor:alternatives}(ii).

\paragraph{(S4) Gate structure.} This condition applies the same rule to the height, so that each equal gain in the surrounding context raises the highest curiosity the item can evoke by an equal amount. Since the height is zero at zero context by Axiom \ref{ax:perceivability} and is set to one at full context, we obtain $A(f) = f$, so the surrounding context itself becomes the height of the curve, as we show in Step 8 of the proof. If each equal gain in the surrounding context closes a constant fraction of the remaining distance to the maximum height, the gate becomes the saturating function $A(f) = (1 - e^{-\beta f})/(1 - e^{-\beta})$ for some $\beta > 0$, which tends to the canonical gate as $\beta \rightarrow 0$, as we show in Corollary \ref{cor:alternatives}(iii).

\subsection{Alternatives to the Canonical Form}\label{app:altproof}

The shape conditions of Theorem \ref{thm:char} select the canonical form from the family of curiosity functions that satisfy the five axioms. We now characterize the alternatives in this family that arise when we relax one condition at a time.

\begin{corollary}[Alternatives]\label{cor:alternatives}
Relaxing the hypotheses of Theorem \ref{thm:char} one at a time yields the following alternatives, each satisfying Axioms \ref{ax:relativity}--\ref{ax:recency} and estimable against the canonical form:

(i) dropping (S2) frees $\psi$ to an arbitrary peaked profile, including in particular the compact-support kernels (Epanechnikov $\psi(y) = \max(0, 1 - y^{2}/\ell^{2})$ and triangular $\psi(y) = \max(0, 1 - |y|/\ell)$, with width $\ell > 0$) and asymmetric profiles;

(ii) replacing the uniform increment of (S3) by a proportional one gives $m(f,\tau) = \tau(1 - f)^{\kappa}$ with $\kappa > 0$, the canonical form being $\kappa = 1$;

(iii) replacing the uniform increments of (S4) by geometric ones gives the saturating gate $A(f) = (1 - e^{-\beta f})/(1 - e^{-\beta})$ with $\beta > 0$, the canonical form being the limit $\beta \rightarrow 0$;

(iv) dropping the $\tau$-independence of $A$ in (S1) \emph{together with} the ceiling of (S4) gives the state-dependent amplitude $A(f,\tau) = a(\tau) f$; dropping the $\tau$-independence alone leaves the canonical form unchanged.
\end{corollary}

Branches (ii) and (iii) are one-parameter families containing the canonical form at $\kappa = 1$ and $\beta \rightarrow 0$, so each yields a nested restriction. Branch (i) is not nested, because the Gaussian is neither a member nor a limit of the Epanechnikov and triangular families, so we compare them by fit across kernels. Such a comparison tests whether curiosity vanishes at a finite gap or decays asymptotically. Branch (iv) is identified only where curiosity is reported directly. Since curiosity utility is $w_{ut} \cdot C_{uit}$ and both $a(\tau_{ut})$ and $w_{ut}$ are indexed by $(u,t)$, a state-dependent amplitude is observationally equivalent to a rescaling of responsiveness in engagement data.

\begin{proof}{Proof}
Each branch removes one hypothesis and re-runs those steps of the proof of Theorem \ref{thm:char} in Appendix \ref{app:charproof} that invoke it; the remaining steps carry over verbatim.

\emph{(i) Dropping (S2).} Steps 1--4 are the only steps invoking (S2), and each lapses: Step 1 derived evenness from it, Step 2 positivity, Step 3 the Cauchy equation, and Step 4 the sign of $c$. Step 0 survives, since it uses only Lemma \ref{lem:wellposed}, the definition of the peak location, and (S1); hence $A(f) > 0$ on $(0,1)$. What still binds $\psi$ is (S1) (continuity, $\psi(0) = 1$, and independence of $(f,\tau)$) together with Proposition \ref{prop:repr}, which requires $F(\cdot,f,\tau)$ to be a peaked profile recentered at $m$; dividing by $A(f) > 0$, $\psi$ must itself be a peaked profile. Conversely, any peaked $\psi$ makes $F = A(f)\,\psi(g - m)$ satisfy Axioms \ref{ax:relativity}--\ref{ax:recency} by Proposition \ref{prop:repr}. The freed family is therefore the set of all peaked profiles. In particular, Step 2's exclusion of zeros lapses, so kernels vanishing identically outside a bounded radius are allowed, and Step 1's evenness lapses, so asymmetric profiles are allowed as well.

\emph{(ii) Proportional increments in (S3).} Steps 1--5 are unaffected, since they invoke (S1), (S2), Axiom \ref{ax:perceivability} and (S3)'s homogeneity clause only, none of which is relaxed. Write $s := 1 - f$ for the remaining ignorance and $\nu(s) := \mu(1 - s)$, so that $\nu$ is increasing by Axiom \ref{ax:optimum}, with $\nu(1^{-}) = 1$ and $\nu(0^{+}) = 0$ by (S3)'s endpoints. Where (S3) requires equal \emph{absolute} steps in $f$ to shift the peak by equal absolute amounts, the proportional counterpart requires equal \emph{proportional} reductions in remaining ignorance to shrink it by equal proportional amounts:
\[
\text{(S3$'$)}\quad \nu(s_{1}s_{2}) = \nu(s_{1})\,\nu(s_{2}),
\quad s_{1},s_{2} \in (0,1).
\]
If $\nu(s_{0}) = 0$ for some $s_{0}$, then $\nu(s_{0}^{1/n})^{n} = 0$ forces $\nu(s_{0}^{1/n}) = 0$ for every $n$; but $s_{0}^{1/n} \rightarrow 1$ and $\nu(1^{-}) = 1$, which gives the same contradiction as in Step 2. Hence $\nu > 0$ and $\chi := \log \circ\, \nu \circ \exp$ is well-defined, monotone, and additive on $(-\infty,0)$, so $\chi(x) = \kappa x$ and $\nu(s) = s^{\kappa}$. Therefore $\mu(f) = (1-f)^{\kappa}$ and $m(f,\tau) = \tau(1-f)^{\kappa}$, where $\kappa > 0$ is the condition for $\mu(1^{-}) = 0$ and for the monotonicity that Proposition \ref{prop:repr} requires: $\partial \gs/\partial f = -\kappa\tau(1-f)^{\kappa-1} < 0$ and $\partial \gs/\partial \tau = (1-f)^{\kappa} > 0$ on $(0,1)$. Step 7's affine $\mu(f) = 1 - f$ is the case $\kappa = 1$.

\emph{(iii) Geometric increments in (S4).} Steps 1--7 are unaffected. (S4) requires $A(f+\delta) - A(f)$ to be independent of $f$, while the saturating counterpart lets successive equal steps close a constant fraction of the remaining headroom,
\[
\text{(S4$'$)}\quad A(f+\delta) - A(f) = \zeta^{\,f}\, \xi(\delta),
\quad \zeta \in (0,1],
\]
for some function $\xi$, which returns (S4) at $\zeta = 1$. Step 8's intercept argument is untouched (it uses only continuity of $F$ and Axiom \ref{ax:perceivability}'s near margin), so $A(0^{+}) = 0$. Letting $f \rightarrow 0^{+}$ in (S4$'$) gives $\xi(\delta) = A(\delta)$, so $A(f+\delta) = A(f) + \zeta^{\,f}A(\delta)$. Exchanging $f$ and $\delta$ and equating the two expressions for $A(f+\delta)$ yields $A(f)\left( 1 - \zeta^{\,\delta} \right) = A(\delta)\left( 1 - \zeta^{\,f} \right)$, so for $\zeta < 1$ the ratio $A(f)/(1 - \zeta^{\,f})$ does not depend on $f$, and the ceiling $A(1^{-}) = 1$ fixes its value at $1/(1-\zeta)$. Writing $\zeta = e^{-\beta}$ with $\beta > 0$,
\[
A(f) = \frac{1 - e^{-\beta f}}{1 - e^{-\beta}},
\]
which is increasing with $A(0^{+}) = 0$, $A(1^{-}) = 1$, and tends to $f$ as $\beta \rightarrow 0$. The canonical linear gate is thus the $\beta \rightarrow 0$ limit of this family.

\emph{(iv) $\tau$-dependent amplitude.} Steps 1--7 never reference $A$'s arguments: (S2) and Axiom \ref{ax:perceivability} determine $\psi$, and (S3) determines $m$. Step 0 gives $A(f,\tau) = F(m(f,\tau),f,\tau) > 0$. Step 8 applies (S4) at each fixed $\tau$, making $A(\cdot,\tau)$ affine with $A(0^{+},\tau) = 0$ from the near margin, and the ceiling $A(1^{-}) = 1$ then gives $A(f,\tau) = f$ for every $\tau$. The $\tau$-independence assumed in (S1) is therefore \emph{recovered}, not needed, and dropping it alone changes nothing. Relaxing the ceiling jointly, to $A(1^{-},\tau) = a(\tau)$ for a continuous $a > 0$, gives $A(f,\tau) = a(\tau) f$ and
\[
F(g,f,\tau) = a(\tau) \cdot f \cdot
\exp\left( -\frac{(g - \tau(1-f))^{2}}{2\sigma^{2}} \right),
\]
which satisfies Axioms \ref{ax:relativity}--\ref{ax:recency} for any such $a$, since a factor depending on $\tau$ alone does not change $\gs$.\Halmos
\end{proof}

\subsection{Verification of the Instantiation}\label{app:verify}

We next verify that the canonical instantiation of Section \ref{sec:measurement}, with the exponentially decayed reference gap, satisfies all five axioms.

\begin{proposition}\label{prop:verify}
The canonical instantiation with decay parameter $\lambda \in (0,1)$ satisfies Axioms \ref{ax:relativity}--\ref{ax:recency}, with unique maximizer $\gs(f,\tau) = \tau(1 - f)$.
\end{proposition}

\begin{proof}{Proof}
Axiom \ref{ax:relativity}: the form depends only on $(g,f,\tau)$. Axiom \ref{ax:recency}: the normalized weights $\lambda^{t-t'}/\sum_{(j,t'') \in \Hut} \lambda^{t-t''}$ of Section \ref{sec:refgap} are non-negative and sum to one, and for $\lambda \in (0,1)$ they are non-increasing and not constant in elapsed time. Axiom \ref{ax:perceivability}: the gate annihilates the expression at $f = 0$; for $f \in (0,1)$ the Gaussian factor vanishes as $g \rightarrow \infty$. The derivative $\partial F/\partial g = -f\sigma^{-2}\left( g - \tau(1 - f) \right) e^{-\left( g - \tau(1 - f) \right)^{2}/2\sigma^{2}}$ has the unique zero $g = \tau(1 - f) \geq 0$, with $\partial^{2}F/\partial g^{2} = -f/\sigma^{2} < 0$ there; hence the maximizer is unique and $\gs = \tau(1 - f)$. Axiom \ref{ax:optimum}: $\partial \gs/\partial f = -\tau < 0$ for all $f \in (0,1)$, $\tau > 0$. Axiom \ref{ax:tracking}: $\partial \gs/\partial\tau = 1 - f > 0$ for all $f \in (0,1)$.\Halmos
\end{proof}

\subsection{Prediction Invariance}\label{app:invariance}

Finally, we prove that the qualitative predictions of Section \ref{sec:canonical} hold for every curiosity function that satisfies the five axioms, so they do not depend on the choice of the canonical form.

\begin{proposition}[Prediction invariance]\label{prop:invariance}
Every curiosity function satisfying Axioms \ref{ax:relativity}--\ref{ax:recency} satisfies:

(i) an interior optimum with dead margins: $\gs(f,\tau) \in (0,\infty)$, $F(g,0,\tau) = 0$, and $F \rightarrow 0$ as $g \rightarrow \infty$;

(ii) leftward peak shift in context: $f < f' \Rightarrow \gs(f',\tau) < \gs(f,\tau)$;

(iii) rightward peak shift in reference gap: $\tau < \tau' \Rightarrow \gs(f,\tau) < \gs(f,\tau')$;

(iv) recency separation: two consumption histories that permute the same gap levels across different times yield different $\tau$, hence different optima, while any model whose reference gap does not depend on recency yields identical optima;

(v) overshoot penalty: $F(g,f,\tau) < F(\gs,f,\tau)$ for all $g > \gs(f,\tau)$, with the loss total in the limit.
\end{proposition}

\begin{proof}{Proof}
(i) Lemma \ref{lem:wellposed}, Corollary \ref{cor:interior}, Axiom \ref{ax:perceivability}. (ii)--(iii) are Axioms \ref{ax:optimum}--\ref{ax:tracking}. (iv) Weights non-increasing and not constant give elapsed times with $\omega_{1} > \omega_{2}$; exchanging consumption events at gaps $g_{a} \neq g_{b}$ across those times changes the Axiom \ref{ax:recency} average by $(\omega_{1} - \omega_{2})(g_{b} - g_{a}) \neq 0$, and Axiom \ref{ax:tracking} converts the change into a shifted optimum. (v) $g > \gs = \max\Gs$ implies $g \notin \Gs$; Axiom \ref{ax:perceivability} gives the limit.\Halmos
\end{proof}

\section{Questionnaire of the Survey-Based Study}\label{app:questionnaire}

This appendix presents the full questionnaire of the survey-based study in Section \ref{sec:labstudy}.

Before the movie recommendations, the questionnaire asked the following background questions, each answered on a seven-star scale unless noted otherwise.

\begin{enumerate}
\item Gender (single choice) and age (numeric).
\item How often do you watch [genre] movies? (1 star = never, 7 stars = very often), asked for each of ten genres, namely action, comedy, drama, science fiction, romance, thriller, animation, horror, documentary, and fantasy.
\item In the past month, to what extent did you watch movies that differ from your usual taste? (1 star = not at all, 7 stars = to a great extent)
\item In the past month, to what extent did you seek out films outside your comfort zone? (1 star = not at all, 7 stars = to a great extent)
\item In the past month, how adventurous was your movie viewing overall? (1 star = not at all adventurous, 7 stars = extremely adventurous)
\item Please name one movie you have already seen and love. (open answer)
\item Any other movies you have already seen and enjoyed? (open answer, optional)
\end{enumerate}

\noindent Items 3--5 form the recent-consumption proxy for the reference gap, and the movie named in item 6 is inserted into the closure question. Figure \ref{fig:questionnaire} shows the beginning of a movie recommendation in the online questionnaire. Each recommendation presents the poster, title, genre, and a short description, followed by six seven-star intensity items, each anchored verbally at both endpoints (1 = not at all, 7 = extremely):

\begin{enumerate}
\item At the current moment, how curious are you about this movie
recommendation? \emph{(curiosity 1)}
\item How much do you want to know more about what this movie is like?
\emph{(curiosity 2)}
\item How strongly does this recommendation make you want to find out more
about it? \emph{(curiosity 3)}
\item How new or unfamiliar is this movie to you? \emph{(perceived novelty)}
\item Given the movies shown so far, how surprising is this recommendation?
\emph{(perceived unexpectedness)}
\item How much do you already know about this movie? \emph{(prior knowledge)}
\end{enumerate}

\noindent followed by a graded consumption item (``Before today, had you seen this film?'': \emph{No, never / Only parts, clips, or the trailer / Yes, the whole film}). After the fifteen recommendations, the closure question inserts the movie that the participant named at the start (``Imagine we recommended another movie chosen to be as similar as possible to [their film]'') and repeats the same six items, and an instructed-response attention check concludes the questionnaire. The order of the movies was randomized for each participant and logged.

\begin{figure}[tbp]
\FIGURE
{\includegraphics[width=0.68\linewidth]{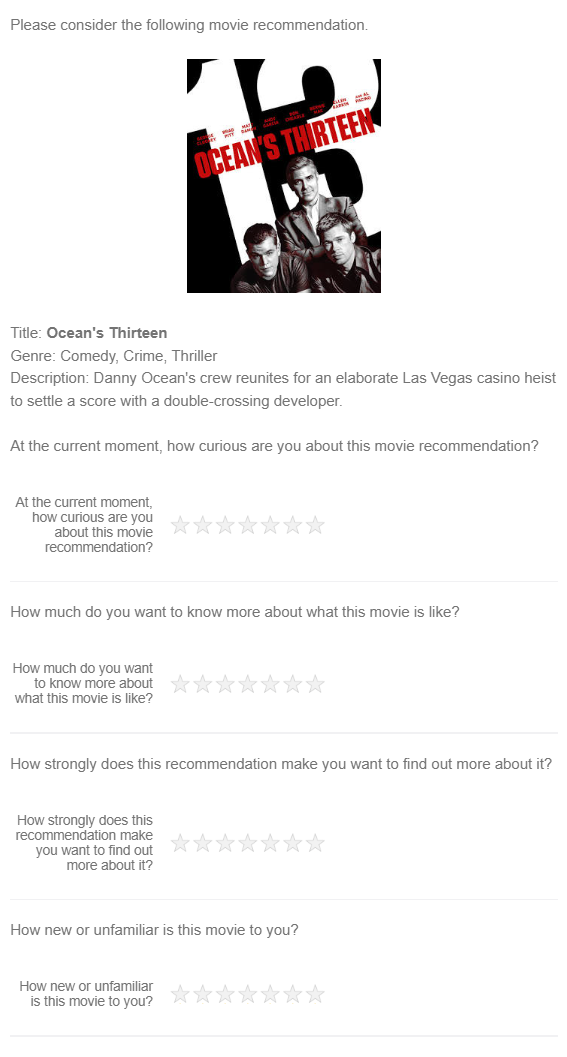}}
{Example of a Movie Recommendation in the Online Questionnaire\label{fig:questionnaire}}
{}
\end{figure}

\section{Implementation of the Curiosity Measure in the Online Controlled Experiment}\label{app:online}

Due to the latency constraint of the production system, we simplify the computation of the three components of the curiosity measure in the online controlled experiment of Section \ref{sec:field}, while we keep the canonical form \eqref{eq:canonical} unchanged. This appendix describes the online implementation and shows that it satisfies all five axioms of Section \ref{sec:axioms}.

\subsection{Online Implementation}

Regarding the information gap, we use the item vectors and consumer states that the production recommender system of the platform already computes, both of which are updated daily. The information gap $g_{uit}$ is the Euclidean distance between the vector of item $i$ and the state of consumer $u$, which is consistent with Section \ref{sec:gap}, where alternative representations of items and consumers can be used under our curiosity framework.

Regarding the surrounding context, let $\mathcal{R}_{ut} \subseteq \Hut$ denote the ten most recent consumption events of consumer $u$ before time $t$. We compute
\[
f_{uit} = \frac{1}{10} \left| \left\{ (j,t') \in \mathcal{R}_{ut} : d(\eb{j},\eb{i}) \leq 0.1 \right\} \right|,
\]
so that $f_{uit}$ is the share of the ten most recently consumed items that lie within a Euclidean distance of 0.1 of item $i$, which replaces the Gaussian kernel of Section \ref{sec:context} with a fixed distance cutoff.

Regarding the reference gap, we use the sliding-window weights of Robustness Check 2 in Section \ref{sec:robustness} with a window length of ten, so that $\tau_{ut}$ is the average information gap of the ten most recently consumed items,
\[
\tau_{ut} = \frac{1}{10} \sum_{(j,t') \in \mathcal{R}_{ut}} g_{ujt'}.
\]

\subsection{Consistency with the Axioms}

The online implementation satisfies all five axioms. Since the canonical form \eqref{eq:canonical} is unchanged, curiosity depends on $(u,i,t)$ only through $(g_{uit}, f_{uit}, \tau_{ut})$, which satisfies Axiom \ref{ax:relativity}, and the gate $f$ and the peak location $\tau(1-f)$ satisfy Axioms \ref{ax:perceivability}--\ref{ax:tracking} for any values of the three components, as we verify in Appendix \ref{app:verify}. In addition, the sliding-window weights assign the weight $1/10$ to each of the ten most recent consumption events and zero weight to all earlier ones, so the weights are non-negative, sum to one, and do not increase with elapsed time, which satisfies Axiom \ref{ax:recency} for every consumer with more than ten consumption events. The online surrounding context also keeps three of the four properties of Section \ref{sec:context}, as it depends only on the consumed items near item $i$, equals zero when none of the ten most recently consumed items lies within the cutoff, and grows with the number of such items, while the fixed cutoff gives every consumed item within the cutoff the same weight.

\end{APPENDICES}

\end{document}